\documentstyle[epsfig,rotating,amssymb]{mn}
\newif\ifAMStwofonts
\ifoldfss
  \ifCUPmtlplainloaded \else
    \NewTextAlphabet{textbfit} {cmbxti10} {}
    \NewTextAlphabet{textbfss} {cmssbx10} {}
    \NewMathAlphabet{mathbfit} {cmbxti10} {} 
    \NewMathAlphabet{mathbfss} {cmssbx10} {} 
  \fi
  \ifAMStwofonts
    \ifCUPmtlplainloaded \else
      \NewSymbolFont{upmath} {eurm10}
      \NewSymbolFont{AMSa} {msam10}
      \NewMathSymbol{\upi}     {0}{upmath}{19}
      \NewMathSymbol{\umu}     {0}{upmath}{16}
      \NewMathSymbol{\upartial}{0}{upmath}{40}
      \NewMathSymbol{\leqslant}{3}{AMSa}{36}
      \NewMathSymbol{\geqslant}{3}{AMSa}{3E}

      \let\leq=\leqslant \let\le=\leqslant
      \let\geq=\geqslant 
    \fi
  \fi
\fi 

\ifnfssone
  \newmathalphabet{\mathit}
  \addtoversion{normal}{\mathit}{cmr}{m}{it}
  \addtoversion{bold}{\mathit}{cmr}{bx}{it}
  \newmathalphabet{\mathbfit} 
  \addtoversion{normal}{\mathbfit}{cmr}{bx}{it}
  \addtoversion{bold}{\mathbfit}{cmr}{bx}{it}
  \newmathalphabet{\mathbfss} 
  \addtoversion{normal}{\mathbfss}{cmss}{bx}{n}
  \addtoversion{bold}{\mathbfss}{cmss}{bx}{n}
  \ifAMStwofonts
    \ifCUPmtlplainloaded \else
      \UseAMStwoboldmath
      \makeatletter
      \new@mathgroup\upmath@group
      \define@mathgroup\mv@normal\upmath@group{eur}{m}{n}
      \define@mathgroup\mv@bold\upmath@group{eur}{b}{n}
      \edef\UPM{\hexnumber\upmath@group}
      \new@mathgroup\amsa@group
      \define@mathgroup\mv@normal\amsa@group{msa}{m}{n}
      \define@mathgroup\mv@bold\amsa@group{msa}{m}{n}
      \edef\AMSa{\hexnumber\amsa@group}
      \makeatother
      \mathchardef\upi="0\UPM19
      \mathchardef\umu="0\UPM16
      \mathchardef\upartial="0\UPM40
      \mathchardef\leqslant="3\AMSa36
      \mathchardef\geqslant="3\AMSa3E

      \let\leq=\leqslant \let\le=\leqslant
      \let\geq=\geqslant 
    \fi
  \fi
\fi 

\ifnfsstwo
  \DeclareMathAlphabet{\mathbfit}{OT1}{cmr}{bx}{it}
  \SetMathAlphabet\mathbfit{bold}{OT1}{cmr}{bx}{it}
  \DeclareMathAlphabet{\mathbfss}{OT1}{cmss}{bx}{n}
  \SetMathAlphabet\mathbfss{bold}{OT1}{cmss}{bx}{n}
  \ifAMStwofonts
    \ifCUPmtlplainloaded \else
      \DeclareSymbolFont{UPM}{U}{eur}{m}{n}
      \SetSymbolFont{UPM}{bold}{U}{eur}{b}{n}
      \DeclareSymbolFont{AMSa}{U}{msa}{m}{n}
      \DeclareMathSymbol{\upi}{0}{UPM}{"19}
      \DeclareMathSymbol{\umu}{0}{UPM}{"16}
      \DeclareMathSymbol{\upartial}{0}{UPM}{"40}
      \DeclareMathSymbol{\leqslant}{3}{AMSa}{"36}
      \DeclareMathSymbol{\geqslant}{3}{AMSa}{"3E}

      \let\leq=\leqslant \let\le=\leqslant
      \let\geq=\geqslant 
    \fi
  \fi
\fi 

\ifCUPmtlplainloaded \else
  \ifAMStwofonts \else 
    \def\upi{\pi}
    \def\umu{\mu}
    \def\upartial{\partial}
  \fi
\fi

\title[Stellar populations in galaxy spheroids.]
{A comparison of stellar populations in galaxy spheroids across a wide
range of Hubble types}
\author[R.N. Proctor \& A.E. Sansom]
  {R.N. Proctor and A.E. Sansom\thanks{E-mail: rproctor@uclan.ac.uk;
aesansom@uclan.ac.uk} \\
   Centre for Astrophysics, University of Central Lancashire,
   Preston, PR1 2HE, UK}

\date{}
\pagerange{\pageref{firstpage}--\pageref{lastpage}}
\pubyear{}

\def\LaTeX{L\kern-.36em\raise.3ex\hbox{a}\kern-.15em
    T\kern-.1667em\lower.7ex\hbox{E}\kern-.125emX}

\begin{document}

\label{firstpage}
\maketitle

\begin{abstract}
We present line-strengths and kinematics from the central regions of 32 
galaxies with Hubble types ranging from E to Sbc. Spectral indices, 
based on the Lick system, are measured in the optical and near infra-red (NIR). 
The 24 indices measured, in conjunction with models of the effects 
of varying abundance ratios, permit the breaking of age/metallicity 
degeneracy and allow estimation of enhancements in specific light elements (particularly 
C and Mg). The large range of Hubble types observed allows direct comparison of 
line-strengths in the centres of early-type galaxies (E and S0) with those 
in spiral bulges, free from systematic differences that have plagued comparisons of 
results from
different studies. Our sample includes field and Virgo cluster galaxies. For early-type
galaxies our data are consistent with previously reported  trends of Mg$_2$ and Mgb with 
velocity dispersion. In spiral bulges we find trends in all indices with velocity dispersion. 
We estimate luminosity-weighted ages, metallicities and heavy element abundance ratios 
(enhancements) from optical indices. These show that bulges are less
enhanced in light ($\alpha$-capture) elements and have lower average age than early-type 
galaxies. Trends involving age and metallicity also differ sharply between early and 
late types. An anti-correlation exists between age and metallicity 
in early types, while, in bulges, metallicity is correlated 
with velocity dispersion. We consider the implications of these
findings for models of the formation of these galaxies. We find that
primordial collapse models of galaxy formation are ruled out by our
observations, while several predictions of hierarchical
clustering (merger) models are confirmed.

\end{abstract}


\section{INTRODUCTION}
\label{intro}
The bulges of spirals, and the spheroidal components of elliptical (E) and S0 galaxies
exhibit many similarities. These include  their colours and  colour
gradients \cite{BP94} as 
well as their morphological and kinematic properties (Bender, Burstein
\& Faber 1993).  The question then 
arises: do these similarities imply similar formation processes? Attempts to answer 
this question using morphology and kinematics have so far failed because the 
observed properties are successfully reproduced by both models proposed
for galaxy formation (i.e. primordial collapse, e.g. Carlberg 1984  and hierarchical 
collapse, e.g. Kauffmann, White \& Guiderdoni 1993 and  Barnes \& Hernquist 1996). On 
the other hand, the use of photometry to constrain the models is severely hampered by 
the degeneracy in colours with respect to age and metallicity (Worthey 1994, hereafter 
W94). Thus, in galaxy populations, the properties above place only weak constraints 
on formation mechanisms and star formation histories (SFHs). 

In order to provide a more sensitive tool for probing composite
populations, W94 used the Lick/IDS spectral features 
\cite{FFBG85}, to estimate pseudo line-strengths (indices), for a range of single age, 
single metallicity stellar populations (SSPs). 
W94 demonstrated that, although individual indices are affected by 
age/metallicity degeneracy, their sensitivities to both age and 
metallicity vary. Consequently, W94 was able to find combinations of
indices that 
broke the degeneracy. Over the last few years, models of SSPs have been
refined and new indices added. Vazdekis et al. (1996; hereafter V96)
and Vazdekis (1999a,b; hereafter V99) 
used more recent (and complete) isochrones for the  SSP calculations, as well as 
including  the MgI and Ca triplet indices in the NIR (Diaz, Terlevich \& Terlevich 1989). 
Higher order Balmer absorption lines (H$_{\delta}$ and H$_{\gamma}$) were 
also included. We shall refer to the collection of indices above as the Lick indices. 

Many studies of early-type galaxies have measured small numbers 
of Lick indices \cite{WFG92,DSP93,Gea97,G97,Tea00a}. These studies show that,
when the commonly observed Mg$_2$ or Mgb indices are plotted against indices 
centred on Fe features (e.g. Fe5227, Fe5335 or their average $<$Fe$>$), 
most early-type galaxies lie outside the range of values covered by
SSPs based on solar metal abundance ratios.
In these galaxies the Mg$_2$ and Mgb index values are significantly 
larger than those of SSPs with similar Fe index values. This illustrates 
the well known Mg excess in early-type galaxies (e.g. Faber 1973; O'Connell 1976; 
Worthey et al. 1992; Henry \& Worthey 1999). 
Indices centred on C features are also seen to lie outside the range 
covered by SSPs \cite{W98,HW99}, implying C is another element whose
relative abundance is enhanced in these populations.  
These studies also show that a positive correlation exists between central 
velocity dispersion and certain Mg indices (e.g. Bender et al.
1993,1998 and Kuntschner 2000; hereafter K00). These trends have been likened to trends in 
colour-magnitude diagrams and interpreted as a metallicity-mass 
relation (i.e. high metallicities are found in high velocity dispersion galaxies due to the 
ability of deep potential wells to retain metals, e.g. Faber 1973;
Kodama et al. 1998). However, this interpretation 
makes no allowance for either the age/metallicity degeneracy in these
indices or their enhancements with respect to other indices. It is also important 
to note that other metallicity-sensitive indices show weak or no
trends with velocity dispersion \cite{FFI96,Tea98,J99}. 

Few studies have included spiral bulges. However, Jablonka, Martin \& Arimoto
\shortcite{JMA96} and Idiart, de Freitas Pacheco \& Costa
\shortcite{IFC96} observed Lick indices in the centres of bulges.
They found correlations of the small number of
indices observed with both bulge luminosity and velocity dispersion.
Goudfrooij, Gorgas \& Jablonka \shortcite{GGJ99} also report
measurements of a small number of indices in bulges and suggest that
the Mg and C excess observed in early types is also present in bulges.
However, in a previous study \cite{PSR00}, we found that a sample of 4 spiral 
bulges lay closer to the solar abundance ratio SSPs, in the
Mg$_2$--$<$Fe$>$ plane, than do early types. Unfortunately, 
none of the bulge studies above were emission corrected, and only
Goudfrooij et al. \shortcite{GGJ99} was fully calibrated to the Lick
system, making full comparisons with early-type galaxies 
difficult. This highlights one of
the advantages of the present study in which such systematic
uncertainties are significantly reduced by the identical
instrumentation and reduction procedures used.

While the sensitivities of Lick indices to abundances of individual
elements, such as Mg and C, complicate their interpretation, they also provide a 
powerful tool for probing 
star formation histories in galaxies, using models of galactic chemical evolution 
(e.g. Vazdekis et al. 1997; Sansom \& Proctor 1998). This is due to 
the difference between production sites of light elements and those of Fe-peak 
elements, i.e. while Mg is mainly produced in type II supernovae (SNII), Fe-peak elements 
are predominantly produced in type Ia supernovae (SNIa). Thus, if reasonable estimates can 
be made of Mg and Fe abundances, constraints can be placed on possible SFHs.
To quantify the effects of individual element abundance enhancements on Lick indices, 
Tripicco \& Bell 1995 (hereafter TB95) modelled the effects of doubling 10 key elements 
in the synthetic spectra of 3 stellar types. They showed that Lick indices centred on C
and Mg features  (e.g. CN$_1$, CN$_2$ or their average $<$CN$>$,
C$_{2}$4668, Mg$_{1}$, 
Mg$_{2}$ and Mgb) are much more sensitive to the abundances of C and Mg than they are 
to overall metallicity ([Z/H]\footnote{We use the standard notation; 
[X/Y]=log[(X/Y)-log(X$_{\odot}$/Y$_{\odot}$)] where X and Y are the masses of either 
individual, or groups of, elements. The term {\bf abundance} is 
used to refer to the mass of a specific element divided by that of H ([X/H]), 
while {\bf abundance ratios} are specified with respect to Fe ([X/Fe]). Overall 
metallicity is represented by [Z/H].}). These indices are then more sensitive to 
the \emph{abundance ratios} of C and Mg (i.e. [C/Fe] and [Mg/Fe]) than to [Z/H]. 
On the other hand, the sensitivities to individual elements of the Fe indices (Fe4383,
Fe4531, Fe5015, Fe5270, Fe5335 or Fe5406) and Ca indices (Ca4227 and 
Ca4455) are comparable to, or less than, their sensitivity
to [Z/H]. It is this range of sensitivities to individual elements and [Z/H]
that permits estimation of ages, metallicities and abundance ratios from index 
values, and thus the constraining of possible SFHs, in galaxy populations.

In this paper we describe the measurement and analysis of 24 Lick indices  
in the centres of 32 galaxies with Hubble types ranging from E
to Sbc. These galaxies were observed in a single observing run, with 
identical instrumental set-ups. We compare the indices and their correlations 
with velocity dispersion in early and late-type galaxies. We
use V99 SSPs and the data of TB95 to estimate luminosity-weighted 
ages, metallicities and abundance ratios in our galaxy sample.
Finally the overall trends in our data are compared to the predictions
of our galactic chemical evolution code for models of galaxy
formation.

In Section \ref{DR} the observations and data reductions are
described, including calibrations and emission corrections.
In Section \ref{results} central values of indices and
kinematics are presented. Comparison of the trends in our data are made
with previous observations. Luminosity-weighted age, metallicity and
abundance ratio
estimates are described in Section \ref{estim}. In Section \ref{comp} 
we discuss the interpretation of our results in terms of composite models of 
possible star formation histories. In Section \ref{concs} we give some
discussion of our results and draw our conclusions. 
Appendix A gives details of velocity dispersion corrections. 
In future papers and in Proctor (PhD thesis - in preparation) 
the spatially resolved  results will be 
reported and the data further exploited to recover 
more detail of the SFHs, using galactic chemical evolution models.

\section{OBSERVATIONS AND DATA REDUCTIONS}
\label{DR}

\begin{table*}
\begin{center}
\caption{WHT Observations. T type,  Hubble type and half-light radius (r$_{e}$) are from de
Vaucouleurs et al. 1991 (hereafter RC3). Central velocity dispersion ($\sigma_{0}$) and 
radial velocity (RV) are the values derived from our blue spectra and are for the central
3.6$\times$1.25 arcsec. Estimated errors are given in brackets (see Section 2.5.1 for details 
of derived kinematics). Distances are mainly from Tully 1988 with the 
exception of more distant galaxies where radial velocities from RC3 were used.
Exposure times and ISIS slit position angle (PA), which is normally along the minor axis
 of each galaxy, are given. Group membership is from Tully 1988.}
\begin{tabular}{|c|c|c|c|c|c|c|c|c|c|}
\hline
Galaxy& T    &Hubble&r$_{e}$&$\sigma_{0}$&RV&Distance&Exposure&PA&Group\\
      & Type &Type  & (")   &(km s$^{-1}$)&(km s$^{-1}$)&(Mpc)&(sec)&(deg)&\\
\hline
NGC2549&-2 & S0 & 17  &143(2)&1076(2) &18.8&3600&87        & Ursa Major cloud  \\
NGC2683& 3 & Sb & 56  &129(2)& 427(2) & 5.7&4800&134       & Leo spur   \\
NGC2832&-4 & E  & 25  &288(5)&6899(6) &91$^{\dag}$&3600&45*&    \\
NGC2831&-5 & E  &     &202(4)&5160(3) &68$^{\dag}$&3600&45*&    \\
NGC3226&-5 & E  & 34  &203(4)&1313(3) &23.4&3600&105       &Leo cloud    \\
NGC3254& 4 & Sbc& 41  &119(2)&1373(2) &23.6&5400&136       &Leo cloud    \\
NGC3301& 0 & S0a& 20  &114(2)&1338(2) &23.3&2400&142       &Leo cloud     \\
NGC3607&-2 & S0 & 43  &240(2)& 930(3) &19.9&2700&30        &Leo cloud     \\
NGC3608&-5 & E  & 34  &208(3)&1219(3) &23.4&2700&165       &Leo cloud     \\
NGC3623& 1 & Sa & 85  &164(2)& 801(2) & 7.3&2400&84        &Leo spur   \\
NGC3769& 3 & Sb &     &46 (9)& 708(3) &17.0&2400&62        &Ursa Major cloud\\
NGC4157& 3 & Sb & 35  & 92(2)& 780(2) &17.0&3600&156       &Ursa Major cloud\\
NGC4192& 2 & Sab& 95  &131(2)&-105(2) &16.8&2700&65        &Virgo cluster  \\
NGC4203&-2 & S0 & 20  &193(3)&1078(2) & 9.7&1200&100       &Coma-Sculptor cloud\\
NGC4216& 3 & Sb & 35  &207(2)& 131(2) &16.8&2400&109       &Virgo cluster \\
NGC4217& 3 & Sb & 55  &95 (4)& 989(3) &17.0&1200&140       &Ursa Major cloud\\
NGC4291&-5 & E  & 17  &292(4)&1701(4) &29.4&2400&20        &CVC cloud     \\
NGC4312&1.5& Sab& 41  &77 (5)& 152(3) &16.8&1200&80        &  Virgo cluster  \\
NGC4313& 2 & Sab&     &69 (2)&1436(2) &16.8&4800&53        &   Virgo cluster  \\
NGC4365&-5 & E  & 50  &254(3)&1221(3) &16.8&2700&130       &   Virgo cluster  \\
NGC4374&-5 & E  & 51  &316(5)&1019(5) &16.8&2400&45        &  Virgo cluster  \\
NGC4419& 1 & Sa & 24  &101(3)&-206(2) &16.8&3600&43        &   Virgo cluster  \\
NGC4526&-2 & S0 & 44  &214(3)& 591(3) &16.8&1500&23        &  Virgo cluster  \\
NGC4552&-4 & E  & 29  &272(4)& 323(4) &16.8&1200&0         &   Virgo cluster  \\
NGC4636&-5 & E  & 89  &243(3)& 919(3) &17.0&2400&60        &   Virgo cluster  \\
NGC4697&-5 & E  & 67  &194(2)& 194(2) &23.3&1200&160       &   Virgo cluster  \\
NGC5322&-5 & E  & 34  &233(3)&1801(3) &31.6&2400&5         &CVC cloud     \\
NGC5354&-2 & S0 & 18  &221(3)&2580(3) &33$^{\dag}$&1200&0  &    \\
NGC5353&-2 & S0 & 15  &280(4)&2230(4) &37.8&1200&0         &CVC cloud     \\
NGC5746& 3 & Sb & 75  &192(2)&1704(2) &24$^{\dag}$&3600&80 &Virgo-Libra cloud      \\
NGC5908& 3 & Sb & 29  &152(2)&3340(2) &29.4&4800&64        &  \\
NGC5987& 3 & Sb & 30  &177(2)&3005(2) &40$^{\dag}$&2400&165&	   \\
\hline	        
\label{data}
\end{tabular}
\end{center}
$^*$ Not along minor axis (Section \ref{obs}).\\
$^{\dag}$ Distances calculated from radial velocities given in RC3 assuming
H$_0$=75 km s$^{-1}$ Mpc$^{-1}$.\\ 
\end{table*}

\subsection{Sample Selection}
The data presented here are from observations made during time awarded
for two separate PPARC PATT proposals. The first was to test the prediction that
young elliptical galaxies should be devoid of hot gas (e.g. Ciotti et al. 1991). 
Bright (B$_T^0 <$ 13) early-type galaxies ($-5 \le T \le -2$), 
with X-ray emission temperatures and luminosities well constrained by ROSAT 
observations, were selected. The second proposal detailed  
an investigation of the stellar populations along the minor axes of bright (B$^{0}_{T}$ $<$
13) spiral bulges ($0 \le T \le 4$). Highly inclined galaxies (inclination $>$ 75$^o$) 
were selected to minimise the effects of disc contamination in the outer regions of
the bulges. Highly inclined galaxies with prominent 
dust lanes covering the centre of the bulge were removed from the
sample, as these would give little information about the bulge centres.
Both studies aimed to use 
the same range of Lick indices to investigate the SFH of galaxies
and to make estimates of luminosity-weighted ages, metallicities and
abundance ratios at a number of points across 
the galaxies, to estimate gradients. 
Fortunately, the two observation runs were scheduled back-to-back allowing 
observations to be made with identical instrument set-ups. The two 
studies result in a sample of 32 galaxies with 24 indices measured (Table
\ref{data}). 
\subsection{Observations}
\label{obs}
Long-slit spectroscopic observations along the minor axes of 11 Es, 
6 S0s and 16 spiral bulges (bulges) were obtained between 1998 February 
28 and 1998 March 3  with the WHT on La Palma. The double beam spectroscope 
(ISIS) was used with a 5700 \AA\, dichroic and Tektronix 1024 square CCDs. On the blue
arm a 300 line/mm grating was used giving a wavelength coverage of 3995-5495 \AA\,
at a dispersion of 1.5 \AA\, per pixel. This range covers 16 indices 
calibrated by Gorgas et al. \shortcite{Gea93} including the extensively 
observed Mg$_{2}$ and $<$Fe$>$ indices, as well as the recently calibrated H$\delta$ 
and H$\gamma$ indices \cite{WO97}. On the red arm a 600 line/mm grating was used, 
giving a wavelength coverage of 8275 \AA\, to 9075 \AA\, at a dispersion of 0.8 
\AA\, per pixel. This range covers the MgI feature at 8807 \AA\, and the  highly 
metallicity-sensitive CaII triplet (Diaz et al. 1989; V96). The
plate-scale was 0.36 
arcsec per pixel on both red and blue arms. The slit, of length 4 arcmin and width 
1.25 arcsec, was placed along the minor axes of the galaxies with the exception of 
NGC 2831 and NGC 2832 which, being in close proximity on the sky, were observed 
simultaneously; i.e. with a 
position angle of 45$^{\circ}$ in both galaxies. A maximum exposure time of 1500 seconds 
was adopted to facilitate cosmic ray removal. Multiple exposures of individual 
galaxies were obtained to achieve the desired signal-to-noise (giving index errors 
of approximately  5\% at r$_{e}$/2). Seeing was $\sim$1.5 arcsec. The total exposure 
time and position angle of the slit are given in Table \ref{data}. Tungsten 
lamp exposures, for flat-fielding, were obtained each night 
on the blue arm. However, due to known fringing effects on the red arm, red 
tungsten lamp exposures were taken just before or after every object exposure, with 
the telescope tracking the object. A total of 5 flux calibration standards and 
24 stars (from Faber et al. 1985) for calibration of the Lick indices were 
observed. A neutral density filter (ND1.8) was used in the stellar
observations. Observations of faint calibration stars and tungsten lamp exposures were
obtained, with and without the neutral density filter, to allow removal
of the filter's spatial and spectral responses. The sample of Lick calibration 
stars was 
selected to possess index values spanning the range of values expected in our 
galaxy sample. The calibration star sample was 
also chosen for good overlap with stars with H$\delta$ and H$\gamma$
measurements reported by Worthey \& Ottaviani \shortcite{WO97}, as well as stars used  in 
the calibration of the CaII index \cite{DTT89}. All calibration stars
possess known heliocentric radial velocities.

\subsection{Basic reductions}
\label{datred}
Unless otherwise stated, data  reductions  were carried  out using the 
CCDPACK, FIGARO and KAPPA packages of Starlink software. Bias removal 
was carried out by the subtraction of an average bias frame,
normalised to the average value in the over-scan region, in each object
frame. After conversion from electrons to photons, variance arrays were 
generated and propagated throughout the reductions.
In the blue, flat-fielding was achieved by division of target frames by the
normalised average of tungsten lamp exposures obtained on the same night. 
However, on the last night suitable tungsten lamp exposures were not obtained.
For this night the flat-field from the first day was used (the day for which 
arc exposures were most similar). Division of target frames by 
tungsten lamp exposures leaves the spectra biased by the smooth spectral 
response of the lamp. This is removed at flux calibration. However, during 
the flat-fielding procedure, features in the blue tungsten lamp spectra were 
identified that moved independently of wavelength calibration. 
The features were in the range 4000 - 4600 \AA\, and were identified with 
features in the dichroic response. The effects of these features are
included in our statistical errors as detailed in
Section \ref{flux}. Flat-fielding of the red data was carried out by division 
of each target frame
by the normalised average of the bracketing tungsten lamp exposures. 
Stellar frames were divided by the normalised neutral density filter
response.
Cosmic rays and bad rows were removed by interpolation across the affected 
areas. Wavelength calibration was carried out by comparison with arc lamp 
exposures taken just before and/or after each exposure. An accuracy of better 
than 0.1 \AA\, was consistently achieved in both red and blue calibrations. 
All object frames were extinction corrected using the extinction 
curve appropriate for La Palma. Flux calibrations derived from multiple observations 
of single stars varied by less than 1.5\% across the region of CCD used,
while those derived from differing stars varied by less than $\sim$ 5\%.
All frames were flux calibrated using the average of the calibration curves 
of 5 flux calibration stars.
Sky estimates were made using the outermost regions of the slit that
were not significantly vignetted. After sky subtraction galaxy frames were 
co-added to form a single frame for each galaxy. The spiral galaxy
NGC 4100 was found to be dominated by 
emission and was excluded from further analysis. Due to the presence of 
telluric absorption lines above 8920 \AA\,, reliable NIR indices could 
not be determined for galaxies with recession velocities above 2200 km s$^{-1}$ 
(i.e. NGC 2831, 2832, 5353, 5354, 5908 and 5987).  

\subsection{Calibrations using stellar spectra}

\begin{table}
\caption{The resolution of the Lick calibrations ($\sigma_{L}$) are given
for reported indices. For all indices, with the exception of G4300, 
offsets represent average differences between the 
published index values and our observations for 24 Lick calibration
stars. For G4300 the offset was modelled as a function of the
observed value (see text). N.B. $<$Fe$>$ is the average of Fe5270 and 
Fe5335 while CaT is the sum of Ca2 and Ca3.}
\begin{tabular}{|l|c|c|c|c|c|}
\hline
Index&Unit&$\sigma_{L}$&Lick  & RMS scatter &RMS error\\
     &    &   (\AA)    &Offset& about offset&per Lick\\
     &    &            &      &             & observation\\      
\hline
H$\delta_A$&\AA& 4.64 &   0.209   & 0.819*   &0.64 \\
H$\delta_F$&\AA& 4.64 &  -0.048   & 0.410*   &0.40 \\
CN$_{1}$& mag &  4.51 &  -0.018   & 0.038*   &0.021 \\
CN$_{2}$& mag &  4.51 &  -0.009   & 0.037*   &0.023 \\
Ca4227  &\AA&    4.34 &   0.027   & 0.189*   &0.27 \\
G4300   &\AA&    4.17 &  \dag     & 0.656*   &0.39 \\
H$\gamma_A$&\AA& 4.04 &   0.561   & 1.014*   &0.48 \\
H$\gamma_F$&\AA& 4.04 &   0.258   & 0.891*   &0.33 \\
Fe4383  &\AA&    3.91 &   1.137   & 1.739*   &0.53 \\
Ca4455  &\AA&    3.87 &   0.226   & 0.454*   &0.25 \\
Fe4531  &\AA&    3.83 &   0.370   & 0.697*   &0.42 \\
C$_{2}$4668&\AA& 3.74 &  -0.276   & 0.437    &0.64 \\
H$\beta$&\AA&    3.61 &  -0.093   & 0.141    &0.22 \\
Fe5015  &\AA&    3.57 &   0.211   & 0.423    &0.46 \\
Mg$_{1}$& mag &  3.57 &   0.007   & 0.006    &0.007\\
Mg$_{2}$& mag &  3.57 &   0.029   & 0.007    &0.008\\
Mgb     &\AA&    3.57 &   0.126   & 0.173    &0.23 \\
Fe5270  &\AA&    3.57 &   0.032   & 0.173    &0.28 \\
Fe5335  &\AA&    3.57 &  -0.040   & 0.238    &0.26 \\
\hspace{3mm}$<$Fe$>$&\AA&    3.57 &-0.004    & 0.147   &0.19 \\
Fe5406  &\AA&    3.57 &   0.053   & 0.129    &0.20 \\
Ca1     &\AA&    3.50 &   0.190   & 0.195    &0.1$^{**}$\\
Ca2     &\AA&    3.50 &   0.208   & 0.226    &0.2$^{**}$\\
Ca3     &\AA&    3.50 &  -0.088   & 0.333    &0.2$^{**}$\\
\hspace{3mm}CaT&\AA&3.50& 0.120   & 0.402    & 0.28\\
MgI     &\AA&    3.50 &   0.031   & 0.093    &0.04$^{**}$\\
\hline
\label{FWHM}
\end{tabular}
$^*$ Features affected by dichroic response.\\
$^{**}$ Typical error given by Diaz et al. (1989) as 5\%.\\
\dag Lick offset=4.34-0.749G4300$_{raw}$ (see Section \ref{flux}).
\end{table}

The original calibrations of stellar line-strengths with photospheric
parameters, upon which the SSPs used here are based, were carried 
out using data from the Lick/IDS scanner \cite{FFBG85}. This instrument 
has a spectral resolution that varies with 
wavelength \cite{WO97}. The Lick data were also not flux calibrated. 
Therefore, in order to calibrate our index measurements to the Lick 
system, it is necessary to compensate for both the differences in 
flux calibration between our data and that in the calibration data, 
as well as the differences in spectral resolution achieved.

\subsubsection{Correction for spectral resolution}
For each index, the value of the Lick spectral resolution 
($\sigma_{L}$ in Table \ref{FWHM}) was estimated from Fig. 7 of Worthey 
\& Ottaviani \shortcite{WO97}, at the wavelength of the mid-point of 
the central band. The spectra of our sample of Lick calibration stars 
were then broadened to the appropriate calibration resolution 
($\sigma_L$), for each index, by convolution with a Gaussian of width 
$\sigma_{B}$ given by:

\begin{equation}
\sigma_{B}^2 = \sigma_{L}^2 - \sigma_{I}^2.
\label{sigB}
\end{equation}

The instrumental broadening for our data ($\sigma_{I}$) was estimated 
from arc lines and found to be 1.5 \AA\, in the blue and 0.7 \AA\, in
the NIR.
After appropriate broadening, stellar indices were evaluated using our
own code. Wavelength range definitions supplied by Worthey on his home 
page were used. For the NIR indices, band definitions and calibration 
resolution ($\sigma_{L}$) were taken from Diaz et al. \shortcite{DTT89} 
(after allowing for a typographical error). Our code was tested using the
stellar data also provided on Worthey's home page for this purpose. Differences 
between the values given by Worthey and those derived by our code from the 
provided spectra were $\leq$ 0.03 \AA\, for  the line features, and $\leq$ 0.002 mag for
molecular band indices. These discrepancies are smaller
than differences caused by re-calibration of  Worthey's data  to our
wavelength resolution and  are probably the result  of
differences   in the   handling   of  partial  bins.

\subsubsection{Flux calibration correction}
\label{flux}
In order to compensate for the differences in flux calibration
between our data and the stellar calibration spectra, the difference 
between measured and published values was calculated, for each index,
in each of the observed calibration stars. For all indices except
G4300, we found no significant correlation between these differences and the measured
values. Therefore, for all indices except G4300, the average difference is 
used as a final correction to the velocity dispersion corrected values. For G4300, the
differences between measured (G4300$_{raw}$) and published values 
exhibited a correlation with G4300$_{raw}$ given by:

\begin{equation}
Offset=4.340-0.749 \times G4300_{raw}.
\label{g4300}
\end{equation}

The final correction to the velocity dispersion corrected values of G4300 
was, therefore, calculated using this equation. However, for one
galaxy (NGC3769), G4300$_{raw}$ lay significantly outside the range of
values covered by our stellar sample. Consequently, the value of G4300
in this galaxy, while reported here, was omitted from further analysis.
Comparison of the measured indices in the 
calibration sample with the published data is shown in Table \ref{FWHM}. 
For blue indices with  band definitions above 4600 \AA, the RMS scatter about 
the offset in our sample is generally dominated by the RMS error associated 
with individual Lick observations. However, the scatter in our data is
significantly greater than this error for indices with band definitions
below 4600 \AA\, (with the exception of Ca4227). This is the wavelength range
affected by the poor removal of the dichroic response identified in Section 
\ref{datred}. For indices in this wavelength range the excess scatter (calculated 
in quadrature) of our data compared to the Lick error has been included in our 
error calculations. This turns out to be a conservative error estimate for
all indices but Ca4227, whose error we assume is
underestimated. For the NIR indices differences between the 
stellar data of Diaz et al. \shortcite{DTT89} and our stellar data were used 
for flux calibration correction. We note that the scatter in our data is 
greater than the RMS error per single observation in the Diaz et al.
\shortcite{DTT89} data (given as typically 5\%). However, these indices
are not used for the purposes of absolute age/metallicity estimates
and this uncertainty has not been included in our errors. For all indices, 
the error in calibration to the 
Lick system was calculated as the standard error in the stellar data,
i.e. $\frac{RMS}{\sqrt{N-1}}$ where N=24 is the number of calibration
stars.

\subsection{Analysis of galaxy spectra}
\label{galspec}
To derive accurate indices from a galaxy spectrum it is first necessary to obtain 
accurate estimates of recession velocity and velocity dispersion from
the spectra. This allows the red-shift to be taken into account and
the indices to be corrected for velocity dispersion (using the 
polynomials detailed in Appendix A). Galaxy data must also be corrected for 
flux calibration (Section \ref{flux}) and emission.

\subsubsection{Measurement of galaxy kinematics}
\label{Kins}

\begin{figure}
\vspace{3cm}
Figure available from \\
http://www.star.uclan.ac.uk/\~{}rnp/research.htm
\vspace{4cm}
\caption{Comparison of measured central velocity dispersions with average
values from Prugniel \& Simien (1997) for 22 galaxies common 
to both studies. The one-to-one line is shown. Prugniel \& Simien (1997) 
error was estimated as the RMS scatter in the individual measurements included in the 
database which include a large range of measurement methods and
apertures. Errors in our data are the standard errors in values derived
from individual calibration stars for each galaxy.}
\label{cfvds}
\end{figure}

Measurements of galaxy kinematics were carried out 
on both the red and blue data using the Fourier quotient technique within 
the IRAF software package. This technique was used as the associated statistical 
errors are significantly less than those associated with the cross-correlation 
technique at the low velocity dispersions typical of spiral bulges. For the 
purposes of velocity dispersion analysis, H$\beta$ and [OIII]5007 emission lines 
were removed by linear interpolation across affected regions in galaxies showing 
strong emission. Velocity dispersion and recession velocity were
estimated using each of the calibration stars. Final
values and errors were taken as the average and standard error in the
individual estimates. Typical errors for velocity dispersion and recession
velocity were found to be $\sim$ 3.1 km $s^{-1}$ and $\sim$ 2.6 km $s^{-1}$
respectively. A comparison of our results for velocity dispersion with 
the average values in Prugniel \& Simien (1997) is shown in Fig. 
\ref{cfvds}. In this figure the average value and RMS scatter in the 
individual measurements quoted in  Prugniel \& Simien (1997)
are taken as the velocity dispersion and its error respectively. 
Despite differences in spectral and spatial sampling, agreement is
reasonably good between the data sets, with a one-to-one line having a reduced 
$\chi^2$ of 1.7. Our galaxy indices were corrected for velocity
dispersions as described in Appendix A.

\subsubsection{Emission Correction}
\label{ecorr}

The H$\beta$, H$\gamma$ and H$\delta$ indices suffer from line-filling 
in galaxies exhibiting emission. Fe5015 is also affected by 
[OIII]5007 emission in such galaxies, while Mgb is affected by 
[NI]5199 emission \cite{GE96}. We estimated [OIII]5007 emission 
in all galaxy spectra in an effort to compensate for the effects of 
emission in H$\beta,\gamma,\delta$ and Fe5015. No attempt was made to 
estimate the (relatively small) corrections to Mgb. Following 
a procedure similar to that used by Gonz\'{a}lez \shortcite{G93}, each 
galaxy spectrum was divided by a series of template spectra. These 
templates were made by red-shifting and broadening each star used 
in the blue kinematic analysis to the measured  galaxy values. The aim of 
dividing the galaxy spectrum by a well matching template is to remove 
common spectral features around the [OIII] line prior to the
measurement of its equivalent width. The band definitions used for our [OIII] line index 
are as follows:\\

\noindent Side band 1: \,\,\,\,4990.0-5001.0 \AA\\
Centre band: \,\,5001.0-5011.0 \AA\\
Side band 2: \,\,\,\,5011.0-5022.0 \AA\\

The average of values derived using each stellar template was taken as
the [OIII] index for each galaxy. The use of the average of a group of 
well matching stellar spectra differs from the method used by Gonz\'{a}lez 
of creating individual templates for each galaxy. Also, [OIII] band 
definitions differ between the two methods.  Gonz\'{a}lez estimated [OIII] 
emission in four galaxies common with our  sample while three of the galaxies 
were observed by Kuntschner et al. \shortcite{Kea01}. Differences in slit 
width and orientation  make direct comparison difficult. However, the results 
are in reasonable agreement ($\sim \pm$ 0.15 \AA). The results of our
OIII estimates are given in Table \ref{bindices}.

Osterbrock \shortcite{O89} shows that the line-strengths of H$\delta$ and 
H$\gamma$ in emission are less than H$\beta$, with 
line-intensities relative to H$\beta$ of approximately 0.25 and 
0.5 respectively across a large range of conditions. However, the continuum 
level in the spectra of all our galaxies show a reduction of approximately 
50\% between H$\beta$ and H$\gamma$. Therefore, using our 
estimates of [OIII] emission, the Gonzalez \shortcite{G93} correlation 
between [OIII] and H$\beta$ emission, and Osterbrock
\shortcite{O89} data for H$\gamma$ and H$\delta$, we applied the 
following corrections:\\

\noindent Fe5015 = Fe5015$_{raw}$+$\Delta$Fe5015,\hspace{5mm}$\Delta$Fe5015 = $-$[OIII]\\
H$\beta$ = H$\beta_{raw}$+$\Delta$H$\beta$,\hspace{21.2mm}$\Delta$H$\beta$ = $-$0.7[OIII]\\
H$\gamma$ = H$\gamma_{raw}$+$\Delta$H$\gamma$,\hspace{21.5mm}$\Delta$H$\gamma$ =$-$0.7[OIII]\\
H$\delta$ = H$\delta_{raw}$+$\Delta$H$\delta$,\hspace{22.5mm}$\Delta$H$\delta$ = $-$0.35[OIII]\\

It can be seen that, due to the reduction in continuum level, H$\gamma$ is as 
sensitive to emission in absolute terms (\AA) as H$\beta$, while the H$\delta$ 
feature is only half as sensitive. However, due to the range of
strengths of both H$\delta$ and H$\gamma$ in SSPs, estimates of age and 
metallicity made using these indices are significantly less affected by
emission than those made using the H$\beta$ index.

An H$\beta$ emission index similar to the [OIII] emission index described
above was also defined as a check on the [OIII] index.
Comparison of [OIII] and H$\beta$ \emph{emission} indices showed that
most galaxies follow the Gonzalez \shortcite{G93} correlation well. However,
six galaxies were noted (all spiral bulges) with significantly aberrant behaviour. 
Three bulges (NGC 4157, NGC 4217 and NGC 4312) show H$\beta$ emission 
substantially greater than that expected from the [OIII], while three
(NGC 3254, NGC 3769 and 
NGC 4313) show strong [OIII], but with no detectable H$\beta$ emission. These 
late-type galaxies do not follow the Gonz\'{a}lez correlation. H$\beta$ values 
for these six galaxies are omitted from our analysis. 

\section{RESULTS}
\label{results}
\subsection{Central Values}
\label{centvals}

\begin{table*}

\vspace{3cm}
Table available from http://www.star.uclan.ac.uk/\~{}rnp/research.htm
\vspace{20cm}
\label{bindices}
\end{table*}

\begin{figure}
\vspace{3cm}
Figure available from \\
http://www.star.uclan.ac.uk/\~{}rnp/research.htm
\vspace{9cm}
\caption{Comparison of our results for early-type galaxies to published 
data from Trager (1998) 
for commonly quoted indices. Also shown as triangles are results
for Fe5270 and Fe5335 from two galaxies from Davies et al. (1993).}
\label{indexcomps}
\end{figure}

\setcounter{table}{3}
\begin{table}
\begin{center}
\caption{NIR central index values (first line) and errors (second
line). All reduction and calibration errors are included 
in the errors. Omitted galaxies are those whose  high
recession velocities redshift one
side-band of these indices into the region of the spectrum affected by
telluric lines.}
\begin{tabular}{rrrrr}
\hline
Galaxy&Ca1&Ca2&Ca3&MgI\\
&(\AA)&(\AA)&(\AA)&(\AA)\\
\hline
NGC 2549&  1.774&  4.528&  3.878&  0.759\\
        &  0.078&  0.076&  0.062&  0.042\\
NGC 2683&  1.697&  4.201&  3.685&  0.665\\
        &  0.101&  0.092&  0.079&  0.057\\
NGC 2832&    -  &    -  &    -  &    -  \\
NGC 2831&    -  &    -  &    -  &    -  \\
NGC 3226&  1.737&  4.242&  3.483&  0.910\\
        &  0.117&  0.110&  0.100&  0.071\\
NGC 3254&  1.687&  4.175&  3.650&  0.671\\
        &  0.116&  0.106&  0.097&  0.068\\
NGC 3301&  1.623&  4.355&  3.775&  0.575\\
        &  0.094&  0.088&  0.074&  0.051\\
NGC 3607&  1.639&  4.169&  3.552&  0.806\\
        &  0.091&  0.089&  0.079&  0.059\\
NGC 3608&  1.797&  4.263&  3.634&  0.854\\
        &  0.100&  0.096&  0.086&  0.060\\
NGC 3623&  1.711&  4.181&  3.658&  0.748\\
        &  0.092&  0.086&  0.072&  0.051\\
NGC 3769&  1.745&  3.965&  3.561&  0.550\\
        &  0.300&  0.268&  0.277&  0.222\\
NGC 4157&  1.573&  4.073&  3.730&  0.543\\
        &  0.188&  0.167&  0.161&  0.131\\
NGC 4192&  1.966&  4.458&  4.170&  0.652\\
        &  0.095&  0.089&  0.075&  0.051\\
NGC 4203&  1.797&  4.276&  3.573&  0.864\\
        &  0.121&  0.111&  0.101&  0.069\\
NGC 4216&  1.794&  4.159&  3.743&  0.732\\
        &  0.086&  0.082&  0.070&  0.046\\
NGC 4217&  2.427&  4.827&  3.877&  0.770\\
        &  0.574&  0.518&  0.576&  0.446\\
NGC 4291&  1.652&  4.197&  3.429&  0.679\\
        &  0.105&  0.104&  0.096&  0.075\\
NGC 4312&  1.530&  3.535&  3.989&  0.344\\
        &  0.596&  0.518&  0.523&  0.331\\
NGC 4313&  1.734&  4.622&  4.145&  0.710\\
        &  0.124&  0.112&  0.102&  0.072\\
NGC 4365&  1.900&  4.430&  3.699&  0.723\\
        &  0.109&  0.105&  0.098&  0.074\\
NGC 4374&  1.740&  4.109&  3.438&  0.597\\
        &  0.104&  0.102&  0.095&  0.075\\
NGC 4419&  1.854&  4.143&  3.813&  0.630\\
        &  0.106&  0.099&  0.088&  0.060\\
NGC 4526&  1.910&  4.311&  3.793&  0.823\\
        &  0.116&  0.108&  0.099&  0.068\\
NGC 4552&  1.603&  3.771&  3.257&  0.705\\
        &  0.111&  0.105&  0.097&  0.074\\
NGC 4636&  1.629&  4.009&  3.539&  0.707\\
        &  0.128&  0.119&  0.111&  0.081\\
NGC 4697&  1.660&  4.343&  3.625&  0.703\\
        &  0.110&  0.102&  0.092&  0.062\\
NGC 5322&  1.858&  4.471&  3.635&  0.778\\
        &  0.110&  0.105&  0.097&  0.072\\
NGC 5354&     -  &    -  &    -  &    - \\
NGC 5353&     -  &    -  &    -  &    - \\
NGC 5746&  1.882&  4.403&  3.852&  0.812\\
        &  0.116&  0.107&  0.096&  0.067\\
NGC 5908&     -  &    -  &    -  &    - \\
NGC 5987&     -  &    -  &    -  &    - \\
\hline
\label{rindices}
\end{tabular}	
\end{center}
\end{table}

Central values for velocity dispersion, given in Table \ref{data}, are those 
derived from the central 3.6$\times$1.25 arcsec of the blue observations with centres
defined as luminosity peaks. Velocity dispersion values derived  using blue spectra are 
greater than those derived from NIR spectra by an average of $\sim$ 
10 km s$^{-1}$ with scatter about this of $\sim$ 20 km s$^{-1}$. However, 
velocity dispersion profiles in some galaxies (e.g. NGC 4192) differ significantly 
between blue and NIR data. This suggests that the two 
wavelength ranges may be detecting differing kinematic populations. 

Central index values and errors for blue indices are given in Table 
\ref{bindices}. Values for Fe5406 in NGC 2831 and NGC 2832
were not determined due to the high recession velocity in these galaxies 
redshifting the blue side-band of this index outside the observed
spectral range. Indices are corrected for both velocity dispersion 
and emission and converted to the Lick system. Reduction and calibration 
errors have all been included in the quoted errors. Also 
included, for indices with band definitions below 4600 \AA, are the
errors due to poor removal of the dichroic response (Section \ref{flux}). 
Comparison of the 
results of our index measurements with galaxies in common with Trager 
\shortcite{T98} are shown in Fig. \ref{indexcomps}. Two galaxies from the 
study of Davies et al. \shortcite{DSP93} are included for the 
Fe indices. The velocity dispersion sensitive Fe5270 and Fe5335 
indices agree within errors. However, for Mg$_2$, an offset of 
$\sim$ 0.008 mag is observed. Such an offset would cause metallicity and 
abundance ratio estimates, made using this index alone, to vary by 
$\sim$ 0.04 dex. Part of the offset is probably the result of the 
slightly wider (1.4 arcsec) and longer (4 arcsec) aperture used in the 
Lick/IDS observations reported in Trager's thesis. Increasing the size 
of the central region from 3.6$\times$1.25 arcsec to 4.5$\times$1.25 arcsec 
(equal surface area) in our measurements results in an average 
reduction in Mg$_2$ of $\sim$ 0.002 mag while Fe5270 
and Fe5335 are both reduced by $\sim$ 0.01 \AA. 
We therefore conclude that, for these indices at least, our data contain 
no significant systematic biases compared to previous observations.

In Table \ref{rindices} values and errors for the fully calibrated NIR 
calcium triplet and MgI features are presented. Values are from the 
central 3.6$\times$1.25 arcsec based on luminosity peaks in the NIR. Quoted errors 
include reduction, calibration and velocity dispersion errors. 
Velocity dispersion correction was based on velocity dispersion values
derived from the NIR galaxy spectra.

\subsection{Correlations between kinematics and indices}

\subsubsection{Early-type galaxies}
\label{early}
\begin{figure}
\vspace{3cm}
Figure available from \\
http://www.star.uclan.ac.uk/\~{}rnp/research.htm
\vspace{3cm}
\caption{Mg$_2$ and Mgb with central velocity dispersion ($\sigma_0$
in km s$^{-1}$).
Circles show our E and S0 galaxies. The thick line shows the mean
trend from K00. Thinner lines show trends from other authors as
detailed in the text. N.B. Both Trager et al. (2000b) and K00
estimated correlations on a log(index) scale, resulting in the
curvatures in these plots.}
\label{plotcorrels}
\end{figure}

\begin{figure*}
\vspace{3cm}
Figure available from http://www.star.uclan.ac.uk/\~{}rnp/research.htm
\vspace{9cm}
\caption{Central metallicity-sensitive indices plotted against log of
central velocity dispersion ($\sigma_0$ in km s$^{-1}$). Thick lines are fits to our
spiral bulge data (solid symbols with symbol size largest for S0/a and
smallest for Sbc). Open symbols show early-type galaxies (circles for Es 
and squares for S0s). Thin lines show trends from K00 for early-type
galaxies. Error bars include all reduction and calibration errors. Uncertainties in
calibration to the Lick system are shown as an isolated error bars in bottom right of
each plot.}
\label{Iwithsig}
\end{figure*}

\begin{table}
\caption{Correlations between indices and velocity dispersion.
The number of galaxies (N) in the fit, slopes, intercepts, 
(un-weighted) correlation coefficients (r) and $\chi^2$ for
best fit correlations are given. Derivation of errors (given in
brackets) are described in the text. For Mg$_2$ and Mgb in early type
galaxies, the correlations of K00 are also given. N.B. For Mgb the K00
correlation has been transformed from magnitudes (as presented in K00) to
Angstrom used in this work. NGC4313 was omitted from correlations for late-type 
galaxies (see Section \ref{late})}
\begin{tabular}{lrrrrr}
\hline
Index&N &dI/d(log $\sigma_0$)&Intercept&r&$\chi^2$\\
\hline
\multicolumn{5}{l}{{\bf EARLY TYPE GALAXIES}}\\      
Ca4227       & 17 &  0.22 (9.36) &  0.76(22.03) &  0.21 &  97 \\
Fe5015       & 17 & -0.50(30.18) &  7.17(71.08) &  0.02 & 105 \\
Fe5270       & 17 & -0.84 (3.13) &  5.18 (7.38) & -0.16 & 108 \\
Fe5335       & 17 & -0.62 (3.10) &  4.32 (7.30) & -0.18 &  66 \\
\hspace{1mm}
$<$Fe$>$     & 17 & -0.72 (2.50) &  4.73 (5.90) & -0.19 & 137 \\
Fe5406       & 15 & -0.38 (1.27) &  2.95 (3.00) & -0.21 &  18 \\
             &	  &              &              &       &     \\
CN$_{1}$     & 17 & 0.086(0.855) & -0.089(2.013)&  0.19 &  21 \\
CN$_{2}$     & 17 & 0.113(0.741) & -0.107(1.745)&  0.23 &  26 \\
\hspace{1mm}
$<$CN$>$     & 17 & 0.099(0.792) & -0.098(1.866)&  0.21 &  47 \\
C$_{2}4668$  & 17 & -1.99(13.45) & 12.73(31.68) & -0.12 & 218 \\
Mg$_{1}$     & 17 & 0.066(0.216) &  0.003(0.510)&  0.35 & 438 \\
Mg$_{2}$     & 17 & 0.114(0.178) &  0.062(0.419)&  0.51 & 392 \\
K00 Mg$_{2}$ & 13 & 0.191(0.023) & -0.127(0.054)&       &     \\
Mgb          & 17 &  2.68 (1.76) & -1.24 (5.39) &  0.61 & 245 \\            
K00 Mgb      & 13 &  2.60 (0.40) & -1.06 (1.36) &       &     \\
             &	  &              &              &       &     \\
H$\delta_{A}$& 17 & -2.11(20.71) &  2.61(48.70) & -0.24 &  29 \\
H$\delta_{F}$& 17 & -1.26 (3.32) &  3.05 (7.80) & -0.43 &  40 \\
H$\gamma_{A}$& 17 & -2.83(44.85) &  1.04(105.53)& -0.19 &  29 \\
H$\gamma_{F}$& 17 & -1.66(31.03) &  2.66(73.04) & -0.21 &  10 \\
H$\beta$     & 17 & -1.74 (0.66) &  5.68 (1.56) & -0.71 &  49 \\
             &	  &              &              &       &     \\
CaT          & 13 & -2.35 (1.94) & 13.31 (4.51) & -0.53 &  52 \\
MgI          & 13 & -0.28 (0.86) &  1.40 (2.00) & -0.51 &  16 \\
\hline       
\multicolumn{5}{l}{{\bf LATE TYPE GALAXIES }}\\         
Ca4227       & 14 &   1.68 (0.22) & -2.44 (0.48) &  0.90 &  21 \\
Fe5015       & 14 &   4.51 (0.34) & -3.77 (0.75) &  0.95 &  21 \\
Fe5270       & 14 &   3.03 (0.14) & -3.46 (0.30) &  0.93 &  19 \\
Fe5335       & 14 &   2.86 (0.57) & -3.50 (1.26) &  0.92 &  35 \\
\hspace{1mm}
$<$Fe$>$     & 14 &   2.95 (0.24) & -3.49 (0.52) &  0.96 &  40 \\
Fe5406       & 14 &   2.09 (0.23) & -2.65 (0.48) &  0.97 &  24 \\
             &    &               &              &       &     \\
CN$_{1}$     & 14 &  0.337(0.088) & -0.687(0.190)&  0.92 &   6 \\
CN$_{2}$     & 14 &  0.347(0.093) & -0.669(0.202)&  0.91 &   8 \\
\hspace{1mm}
$<$CN$>$     & 14 &   0.340(0.089)& -0.683(0.193)&  0.91 &  10 \\

C$_{2}4668$  & 14 &  11.80 (1.53) &-18.78 (3.39) &  0.96 &  84 \\
Mg$_{1}$     & 14 &  0.230(0.042) & -0.377(0.091)&  0.94 & 155 \\
Mg$_{2}$     & 14 &  0.413(0.027) & -0.623(0.060)&  0.97 & 170 \\
Mgb          & 14 &   5.37 (0.22) & -7.43 (0.48) &  0.98 &  70 \\
             &	  &               &              &       &     \\
H$\delta_{A}$& 14 & -12.34 (1.93) & 25.72 (4.18) & -0.93 &  24 \\
H$\delta_{F}$& 14 &  -6.04 (0.08) & 13.80 (0.17) & -0.95 &  38 \\
H$\gamma_{A}$& 14 & -14.33 (2.00) & 26.41 (4.40) & -0.88 &  24 \\
H$\gamma_{F}$& 14 &  -7.04 (1.63) & 14.68 (3.59) & -0.84 &  10 \\
H$\beta$     &  9 &  -1.44 (0.74) &  5.09 (1.64) & -0.81 &  16 \\
             &    &               &              &       &     \\
CaT          & 12 &  -1.03 (5.88) & 10.25(12.06) &  0.28 &  50 \\
MgI          & 12 &   0.39 (0.31) & -0.12 (0.63) &  0.78 &   9 \\
\hline
\label{correls}
\end{tabular}
\end{table}

\begin{figure}
\vspace{3cm}
Figure available from \\
http://www.star.uclan.ac.uk/\~{}rnp/research.htm
\vspace{19cm}
\caption{Age sensitive indices against log of central velocity dispersion. Symbols
and errors as Fig. 4.}
\label{AgewithSig}
\end{figure}

\begin{figure}
\vspace{3cm}
Figure available from \\
http://www.star.uclan.ac.uk/\~{}rnp/research.htm
\vspace{8cm}
\caption{Central indices with Hubble (T) type. Open symbols are
early types and filled symbols are late types.}
\label{Htype}
\end{figure}

Correlations between indices sensitive to $\alpha$-elements (e.g. Mg$_2$ and Mgb) and 
velocity dispersion have been widely observed in early-type galaxies 
\cite{BBF93,G93,J97,Ber98,Cea00,K00,Tea00b}. Fig. \ref{plotcorrels} 
summarises results for Mg$_2$ and Mgb from these studies. Although there 
is generally good agreement for the slopes of the correlations, there are 
offsets between studies. Such offsets may result from differences in calibration 
to the Lick system, differences in aperture size and/or orientation and systematic 
differences in velocity dispersion estimates. Values of indices in the 
early-type galaxies of our sample are also shown in Fig. \ref{plotcorrels}. 
Correlations from the fully calibrated K00 study of Fornax cluster galaxies
(shown as thick lines in Fig. \ref{plotcorrels}), most closely match our results, 
although our sample covers a significantly narrower range of velocity
dispersion than the K00 sample (a result of 
our selection of \emph{bright}, nearby ellipticals).\\

In Fig. \ref{Iwithsig} we show plots of selected metallicity-sensitive
indices from the blue spectra against log velocity dispersion for our galaxy 
sample. Indices omitted are those 
most severely affected by the dichroic response problem outlined in  Sections 
\ref{datred} and \ref{flux}. Table \ref{correls} details the fits (minimising $\chi^2$) 
of indices 
against velocity dispersion for both early and late-type galaxies in our sample. 
Values given are from fits with y-axis errors only (which dominate in most cases). 
Errors in slope and intercept are taken as half the difference between the fit 
using index errors for $\chi^2$ minimisation and that using velocity dispersion 
errors. From Table \ref{correls} it can be seen that in the
metallicity sensitive indices of early-type galaxy 
sample we detect no significant (3$\sigma$) slopes. 
However, our results for Mg$_2$ and Mgb are consistent with the positive trends with velocity 
dispersion noted by previous studies (Fig. \ref{plotcorrels}). If we compare the 
scatter of our Mg$_2$ and 
Mgb data about our best fit lines (0.02 mag and 0.3 \AA\, respectively) with their 
scatter about the K00 correlations (Fig. \ref{plotcorrels}), we find no significant 
difference. We tested the fit of the K00 correlations to our early-type galaxy 
data for all common indices. For the majority of indices our scatter 
about the K00 correlations is similar to scatter of the K00 data. Consequently, in Fig. 
\ref{Iwithsig}  we use K00 correlations (shown as thin lines) for all common indices 
for purposes of comparison to bulges. Most previous authors report weak or no trends 
in Fe indices (Fisher el al. 1996; Trager et al. 1998; 
J{\o}rgensen 1999), with velocity dispersion for early-type galaxies. Within the 
narrow velocity dispersion range covered by our sample, we echo this finding (Fig. 
\ref{Iwithsig} and Table \ref{correls}). However, K00 found correlations for all the 
metallicity-sensitive indices he observed. We note the possibility that such correlations 
would have been present in our data had the velocity dispersion range of our study been 
larger. For Fe5406 we find values consistently above the K00 correlation. However, 
the side bands of this index lie in a region of our spectra where vignetting effects 
are beginning to appear. Consequently, it is possible that a 
small undetected systematic effect may be biasing this index in our results. Previous 
authors  have reported that the calcium indices (Ca4227 and Ca4455) follow similar 
trends to Fe indices, rather than the trends in enhanced indices such
as Mg$_2$ \cite{Vea97,Tea98}. We again echo this finding within the narrow range 
of velocity dispersions of our sample. 

Plots of age sensitive indices with velocity dispersion are shown in Fig. 
\ref{AgewithSig}. In early-type galaxies no significant trends with velocity 
dispersion are evident in the H$\delta$ and H$\gamma$ indices. However, for 
the H$\beta$ index a  $\sim$3$\sigma$ correlation is found (Fig. \ref{AgewithSig} 
and Table \ref{correls}). The slope of this weak correlation is
steeper than 
that found by K00 (dH$\beta$/d(log $\sigma$)$\sim$ -0.4) which is shown
as a thin line in Fig. \ref{AgewithSig}. 
The presence of such a correlation in the highly age sensitive H$\beta$
index, while the more degenerate H$\delta$ and H$\gamma$ indices show no
correlations, suggests the presence of a trend of increasing
age and/or decreasing metallicity with velocity dispersion in the
early-type galaxies.

S0 galaxies, shown as open squares in Fig. \ref{Iwithsig}, follow similar trends 
to elliptical galaxies in all indices. The similarity in indices in E
and S0 types can also be seen in Fig. \ref{Htype} 
where the values of key indices are compared for all Hubble types.\\
In this plot indices show no correlation with Hubble type for spiral
bulges.

\subsubsection{Late-type galaxies}
\label{late}
All metallicity-sensitive indices in the blue spectra of our late-type
galaxy sample show strong correlations with 
velocity dispersion (Fig. \ref{Iwithsig}, Table \ref{correls}), including those not 
shown in Fig. \ref{Iwithsig}. One outlier to these trends for late-type galaxies is 
NGC 4313, which exhibits a strong central depression in log $\sigma$. This galaxy has 
been omitted from the line fitting procedure. The highest velocity dispersion bulges 
show index values coincident with those of early-type galaxies. Indeed, for all indices 
in our sample, early and late types form a continuous locus in the index-velocity
dispersion plane. However, the slopes of the correlations in late-type galaxies are 
significantly steeper than those found by K00 for early-type galaxies (Fig. \ref{Iwithsig}). 
In general, the scatter about the correlations of late-type galaxies is also smaller than 
that of early types. Anti-correlations with velocity dispersion are also evident among 
the late-type galaxies in all age sensitive indices (Fig. \ref{AgewithSig}) with low velocity 
dispersion bulges having stronger hydrogen absorption lines (aberrant  emission galaxies 
detailed in Section \ref{ecorr} are omitted from line fitting in the
case of H$\beta$). The strong trends of both 
metallicity and age sensitive indices with velocity dispersion in late-type galaxies suggests 
that SFH is closely associated with the depth of potential well in the centres of bulges. 

\subsubsection{Red data}
\begin{figure}
\vspace{3cm}
Figure available from \\
http://www.star.uclan.ac.uk/\~{}rnp/research.htm
\vspace{8cm}
\caption{CaT and MgI against central velocity dispersion. Symbols and errors as 
Fig. 4}
\label{RedwithSig}
\end{figure}

CaT (the sum of CaII at 8498 \AA\, and 8542 \AA) and MgI 
are plotted against velocity dispersion in Fig. \ref{RedwithSig}. Trends with 
velocity dispersion are not as clear in these indices. In early-type galaxies,  
the  MgI index shows behaviour with velocity dispersion different to that of the 
Mg$_1$, Mg$_2$ and Mgb indices in the blue. However, it is difficult to make 
direct comparisons because, as previously suggested (Section \ref{centvals}), 
NIR indices may be sampling different stellar populations to those
sampled by blue indices.

\subsubsection{Potential disc contamination of late-type galaxies}
Late and early-type galaxies in our sample exhibit  similar index values in the 
region where their velocity dispersions overlap (150 - 200 km s$^{-1}$). However, 
towards lower dispersions, bulges lie systematically further from the K00 
trends for early-type galaxies (Fig. \ref{Iwithsig}). It was a concern that this 
behaviour may be the result of disc contamination. Khosroshahi, Wadadekar 
\& Kembhavi \shortcite{KWK00} carried out bulge-disc decomposition of a 
number of edge-on (i$>$50$^o$) spiral galaxies. They show that for Hubble 
types earlier than Sbc (T$<$4) the bulge-to-disc central luminosity ratio 
always exceeds 10; i.e. no more than 10\% of the observed light originates 
in the disc. They also show that this ratio increases to $\sim$ 1000 in the 
case of Sa galaxies. The worst-case assumption for metallicity-sensitive 
indices is that of a smooth continuum contribution from the disc. Under such
circumstances, a maximum reduction of $\sim$ 10\% is expected for both line and 
molecular band indices. We compare the K00 correlations for metallicity-sensitive 
indices in early-type galaxies with our correlations for late types in  Table \ref{cfEL}.
Values of indices on the correlations are given, for our bulge data and the K00 
correlations, at log $\sigma_0$=2.0 (mid-range value for bulges). Percentage
differences between the two correlations can be seen to vary from -27\%
to +5\%. Clearly the difference can not be modelled by the simple addition of 
continuum to the bulge light. It should also be noted that, for the Mg indices, a minimum bulge 
contribution of $\sim$ 25\% is required to account for the differences between 
correlations. This is well in excess of the 10\% maximum expected from Khosroshahi 
et al.\shortcite{KWK00}, and the difference increases at velocity dispersions 
less than 100 km sec$^{-1}$. We also note that there is no evidence that the 
indices of late-type galaxies correlate with Hubble type (Fig. \ref{Htype}) as 
might be expected if significant disc contamination were present. 

We therefore 
conclude that while metallicity-sensitive indices could be depressed by as much 
as 10\% by disc contamination, this effect is not evident in our data, nor can it 
explain the observed differences between correlations in early and late-type galaxies.

\begin{table}
\begin{center}
\caption{Index values from  correlations at log($\sigma_0$)=2.0. The
percentage difference between the values from the two correlations is
also given.}
\begin{tabular}{lrrr}
\hline
Index&K00&Spirals&\%\\
\hline
C$_2$4668& 5.37 & 4.81 & -10\% \\
Fe5015   & 5.02 & 5.25 &  +5\% \\
Mg$_1$   & 0.114& 0.083& -27\% \\
Mg$_2$   & 0.255& 0.194& -24\% \\
Mgb      & 4.14 & 3.31 & -20\% \\
Fe5406   & 1.69 & 1.52 & -10\% \\
$<$Fe$>$ & 2.69 & 2.40 & -10\% \\
\hline
\label{cfEL}
\end{tabular}
\end{center}
\end{table}

\subsection{Diagnostic index plots}
\label{diagnostics}

\begin{figure*}
\vspace{3cm}
Figure available from http://www.star.uclan.ac.uk/\~{}rnp/research.htm
\vspace{3cm}
\caption{$<$Fe$>$ vs Mg$_2$ (left) for our WHT data and three galaxies
from Proctor et al. 2000 (solid squares). Values for fourth galaxy from
Proctor et al. 2000 (NGC 3623) agree (within errors) with values from 
the WHT observation. All other symbols are as Fig. 4. Grid lines show
V99 solar abundance ratio SSPs at ages of 1.5, 2, 3, 5, 8, 12 and 17 Gyr and
[Fe/H]$_{SSP}$ of $-$1.7, -1.0, -0.5, -0.25, 0.0, +0.25, and +0.5
dex. In the left plot the dashed line shows the correlation of Gorgas et al.
(1997) for early-type galaxies, which is consistent with our data. 
The thick solid line shows the range of index values covered by the
primordial collapse models described in Section \ref{pc}, while the
short, thinner solid line shows the range of values covered by the merger models in
Section \ref{merger}. Error bars in the bottom right of each plot refer
to errors incurred in calibrating onto the Lick system.}
\label{diags}
\end{figure*}

\begin{figure*}
\vspace{3cm}
Figure available from http://www.star.uclan.ac.uk/\~{}rnp/research.htm
\vspace{20cm}
\caption{Metallicity-sensitive indices plotted against H$\beta$.
Symbols as Fig 4. Fe indices are shown on the left, while indices
sensitive to Mg and C abundance ratios are shown on the right.
Uncertainty in calibration to the Lick system is
shown top right of each plot. Short, thick and thinner lines show
primordial collapse and merger models respectively. These and the grid
lines are as in Fig. \ref{diags}.}
\label{IvsHbeta}
\end{figure*}

\begin{figure}
\vspace{3cm}
Figure available from \\
http://www.star.uclan.ac.uk/\~{}rnp/research.htm
\vspace{14cm}
\caption{Calcium sensitive indices; Ca4227 (top) and NIR CaT (middle)
plotted against H$\beta$. Also shown is the MgI index (bottom) against H$\beta$. 
Uncertainty in calibration 
to the Lick system is shown bottom right of each plot. Symbols as Fig. 4.}
\label{CavsHbeta}
\end{figure}

In this section we compare the results of our index determinations to the 
values predicted for SSPs. We use V99 SSPs for our analysis. These
(web published) SSPs are based on Bertelli et al. \shortcite{BBCNF}
isochrones (as detailed in V96), and  Vazdekis \shortcite{V99a}
transformations to the observational plane. We use V99 SSPs rather
than those of W94 and Worthey \& Ottaviani \shortcite{WO97}, 
as V99 make use of the more up-to-date
and complete isochrones of Bertelli et al. \shortcite{BBCNF}.  
V99 also includes the CaT and MgI indices in their 
calculations of SSP index values. Metallicities derived from degeneracy breaking 
diagnostic plots using V99 SSPs are generally higher by $\sim$ 0.1 dex 
than those implied by W94 SSPs. For galaxies older than 
$\sim$ 5 Gyr, V99 SSPs also imply ages younger by $\sim$ 0.15 dex. 

The $<$Fe$>$ vs Mg$_2$ diagnostic plot is shown in Fig.
\ref{diags}. This plot clearly shows the enhancement of the Mg$_2$
index (or, equivalently, the deficiency in the
$<$Fe$>$ index) in early-type galaxies with respect to solar abundance 
ratio SSPs. The plot also demonstrates the age/metallicity
degeneracy of these two indices. The grid includes SSPs ranging
from 1.5 to 17 Gyr in age and with metallicities ([Fe/H]$_{SSP}$)
covering the range [Fe/H]$_{SSP}$=-1.7 to +0.5. The good agreement between 
our data and the correlation of Gorgas et al. 
\shortcite{Gea97} for early-type galaxies is also illustrated. 
Late-type galaxies seem to span the SSP grid. This echos the tentative finding 
of our previous study (Proctor et al. 2000), 
that late-type galaxies exhibit abundance ratios closer to solar than
those found in elliptical galaxies. The 4 bulges observed in Proctor et al.
\shortcite{PSR00} were not fully calibrated to the Lick system due to a lack of 
observations of suitable calibration stars. However, we note that most fully 
calibrated studies find that small positive corrections are required to compensate Mg$_2$ for 
the lack of flux calibration in the Lick observations, while Fe5270 and Fe5335 
require only very small corrections (e.g. Worthey \& Ottaviani 1997; K00). The 
Lick offsets given in Table \ref{FWHM} for the present data set were thus applied 
to the Proctor et al. \shortcite{PSR00} data. The Sa galaxy NGC 3623 was observed 
in both studies. The two independent values for both Mg$_2$ and $<$Fe$>$ in this
galaxy were within errors. Consequently, this galaxy was not included
when the results of the Palomar study were added to Fig. \ref{diags}
(solid squares). Additional errors of 0.01 mag and 0.04 \AA\, have
been added in quadrature to Proctor et al. \shortcite{PSR00} 
errors for Mg$_2$ and $<$Fe$>$ respectively to allow for uncertainty
in the corrections to the Lick system.\\

Given the tendency, reported by
previous authors, for Ca indices to follow similar
trends to Fe indices \cite{Vea97,Tea98}, the NIR CaT and
MgI indices allow plotting of a diagnostic plot similar to $<$Fe$>-$Mg$_2$ in the blue. 
This diagnostic plot (CaT$-$MgI) is also shown in Fig.
\ref{diags}. In this figure, early-type galaxies lie
within with the SSP grid. A simplistic interpretation would be
that this indicates [Mg/Ca]=0 and, assuming that Ca follows the
same trends as Fe, a value [Mg/Fe]=0. However, 
when these indices are plotted against age sensitive indices (e.g. H$\beta$) 
problems with the MgI index become apparent (Section \ref{break}) and
this interpretation can not be sustained.

\subsubsection{Breaking the degeneracy}
\label{break}
Plots that most clearly break the age/metallicity degeneracy are
shown in Figs \ref{IvsHbeta} and \ref{CavsHbeta}. 
These plots show metallicity-sensitive indices against H$\beta$ 
and compare galaxy values to V99 SSP predictions. We show indices 
against H$\beta$ since it is the most age sensitive index and other 
age sensitive indices (H$\delta_{A,F}$, G4300 and  H$\gamma_{A,F}$) are 
among those effected by the dichroic response problem in our data. 

The H$\delta$ and H$\gamma$ indices were also not among the indices
whose sensitivities to element abundance ratios were modelled by TB95. 
Fig. \ref{IvsHbeta} shows indices sensitive to Fe 
on the left, while indices sensitive to abundance ratios are shown on the 
right. Early-type galaxies in Fig. \ref{IvsHbeta} exhibit
different trends in Fe and abundance-ratio-sensitive indices. In these galaxies, 
the \emph{increasing} H$\beta$ with line strength in Fe indices contrasts 
with the \emph{decreasing} H$\beta$ with line strength in 
abundance ratio sensitive indices. 
The trend in Fe indices suggests that Fe abundance is anti-correlated with 
age in early-type galaxies, while the growing disparity between Fe and
abundance ratio sensitive indices suggests increasing abundance enhancement in light
elements with age.

In late-type galaxies, H$\beta$ \emph{decreases} with all metallicity-sensitive 
indices in the blue (Fig. \ref{IvsHbeta}). The position of these galaxies 
with respect to the SSP grids is also more consistent between
plots of Fe indices and those sensitive to abundance ratios. 
This reflects abundance ratios closer to solar in these objects, as noted in
Proctor et al. 2000. The positions of late-type galaxies also suggests that the
populations are relatively young (luminosity-weighted ages $\le$ 5
Gyr) with a wide range of metallicities.

Fig. \ref{CavsHbeta} shows a plot with H$\beta$ of the blue Ca4227
(top plot). Comparison of this plot with Fig. \ref{IvsHbeta}
shows the similarity in behaviour of  Ca and Fe indices previously
reported \cite{Vea97,Tea98} with Ca4227 suggesting
lower metallicities at increased age. Also shown in Fig. \ref{CavsHbeta} are the
NIR CaT and MgI indices plotted against H$\beta$. A note of caution here is that we are
comparing indices derived from NIR spectra with H$\beta$ derived from
the blue spectra. As previously noted, we suspect that the two
wavelength ranges may be sampling differing populations. However,
under the assumption that the differences in age and metallicity of
the populations sampled by NIR and blue wavelength bands are
reasonably small and uniform across our galaxy sample, we attempt
to interpret these diagrams.

The CaT index again shows the similarity in behaviour of Ca and Fe
indices. That is, the trend of increasing H$\beta$ with CaT, 
is similar to those exhibited by blue Fe indices. However, in the MgI-H$\beta$ plot, 
while some galaxies retain the behaviour of blue Mg indices, many galaxies are 
displaced to extremely low metallicities when compared to the 
SSP grids. These galaxies tend to be both the oldest, and the highest
velocity dispersion. However, MgI is a weak index, prone to measurement 
problems, therefore further observations of MgI are needed to ascertain 
its behaviour relative to SSPs. We therefore draw no strong conclusions from
either the MgI$-$H$\beta$ or MgI$-$CaT plots.

\section{ESTIMATION OF AGE, METALLICITY AND ABUNDANCE RATIO ENHANCEMENT}
\label{estim}
Our aim is to use the indices of solar-neighbourhood abundance ratio SSPs 
as the basis for estimating  luminosity-weighted ages, metallicities
and abundance ratios of galaxies. This
requires us to estimate the effects on indices of the non-solar abundance 
ratios observed in many galaxies (e.g. Davies et al. 1993). Thus we aim to construct grids of 
SSPs with varying age, [Fe/H]$_{SSP}$ \emph{and} abundance ratio for comparison 
with observations. This requires knowledge/estimation of:\\

\begin{itemize}
\item The abundance ratio pattern in the local stars used to construct SSPs.\\

\item The difference between the local abundance ratio (SSP) pattern and that in the 
galaxy populations being studied.\\

\item The effects such a difference would have on photospheric conditions 
(surface gravity (log g), effective temperature (T$_{eff}$) and luminosity) 
in whole populations i.e. the effects on isochrones.\\

\item The effects of differences in abundance ratios on the strength of 
individual indices in stars, assuming fixed photospheric conditions.\\
\end{itemize}

Armed with the above it is possible to estimate corrections to the indices
of solar abundance ratio SSPs for non-solar abundance ratios, for comparison 
to observations. The above points are discussed in turn below.

\subsection{The local abundance pattern}
\label{lap}
Stellar metallicity estimates, used in the construction of both W94 and 
V99 SSPs, were based on analysis of solar-neighbourhood stars (e.g. Edvardsson et al.
1993 in V99 and Hansen \& Kj{\ae}rgaard 1971; 
Gustafsson, Kj{\ae}rgaard \& Andersen 1974 in W94). The wavelength ranges used generally 
contain a large number of Fe lines, thus providing good estimates of iron 
abundance ([Fe/H]). However, solar-neighbourhood stars possess considerable 
scatter in the abundance ratios of individual elements. In high metallicity 
stars ([Fe/H] $\geq$ solar) abundance ratios show moderate scatter about solar 
values \cite{Eea93,FG98}. Thus for SSPs, the calibration of which averages a 
large number of local stars, we may assume [Z/H]$_{SSP}$=[Fe/H]$_{SSP}$ for 
[Fe/H]$_{SSP}$$\geq$0. On the other hand, solar-neighbourhood stars with 
[Fe/H] $<$ 0 exhibit non-solar abundance patterns, with $\alpha$-elements 
(e.g. O, Ne, Na, Mg, Al, Si, S, Ar, Ca) enhanced with respect 
to Fe peak elements (e.g. Fe, Ni, Cr) \cite{Rea91,Eea93,FG98,IT00}. The
$\alpha$-element abundances are also enhanced with respect to C, which has solar 
abundance ratio (or below) down to very low metallicities (Ryan et al. 1991). 
Abundance ratios of $\alpha$-elements ([$\alpha$/Fe]) in these stars
increase linearly from [$\alpha$/Fe]$\sim$0.0 at [Fe/H]=0.0 to 
[$\alpha$/Fe]$\sim$+0.3-0.5 (depending on the element concerned) at [Fe/H]$=-1$.
Moderate scatter in the abundance ratios of individual elements
about these trends is again observed. As SSPs with [Fe/H]$_{SSP}$ $<$ 0 are
therefore based on stars with non-solar abundance ratios, we assume that [Fe/H]$_{SSP}$ 
more closely approximates [Fe/H] in the stars than [Z/H]. Thus, in low metallicity
SSPs (and any other non-solar abundance ratio population) we must assume a 
relationship between [Z/H], [Fe/H] and [E/Fe] (where E is the mass of 
all elements with enhanced abundance ratios - see Table \ref{abs},
column 3). As non-linear effects are small \cite{TCB98}, and following
Trager et al. (2000a; hereafter T00a), we assume a linear relationship:

\begin{equation}
[Z/H]=[Fe/H]+A[E/Fe].
\label{tea1a}
\end{equation}

\begin{table}
\begin{center}
\caption{Mass of individual elements as a fraction of total metals
(X$_i$/Z)$_{\odot}$ in the Sun. Data is from Cox (2000). The total mass fraction
of metals in the Sun (Z$_{\odot}$) is assumed to be 0.0189. Elements
enhanced in low-metallicity, solar-neighbourhood stars and in galaxy
populations are identified by a +. These are the elements which are 
included in the enhanced group (E) in each case. The final column 
indicates (with Y) the elements modelled by TB95. The last row in the
second
column shows the total fraction of all tabulated elements in the Sun.
For the other 3 columns totals indicate solar proportions of elements 
identified by + or Y. }
\begin{tabular}{lrccc}
\hline
Element&(X$_i$/Z)$_{\odot}$&Low [Fe/H]& Galaxy&Modelled\\
&&stars&populations&by TB95\\
\hline
C   & 0.1619 &&+&Y\\
N   & 0.0583 &&+&Y\\
O   & 0.5054 &+&+&Y\\
Ne  & 0.0921 &+&+&\\
Na  & 0.0018 &+&+&Y\\
Mg  & 0.0343 &+&+&Y\\
Al  & 0.0030 &+&+&\\
Si  & 0.0370 &+&+&Y\\
S   & 0.0193 &+&+&\\
Ar  & 0.0054 &+&+&\\
Ca  & 0.0034 &+&&Y\\
Cr  & 0.0009 &&&Y\\
Fe  & 0.0719 &&&Y\\
Ni  & 0.0039 &&&\\
\hline
Total&0.9986&0.7017&0.9185&0.8749\\
\hline
\label{abs}
\end{tabular}
\end{center}
\end{table}

\noindent The differential form of which:

\begin{equation}
\Delta[Z/H]=\Delta[Fe/H]+A\Delta[E/Fe] 
\label{tea1d}
\end{equation}

\noindent may be used to derive estimates of the factor $A$ in low metallicity 
stars by comparing a solar composition with one in
which all $\alpha$-elements are doubled in abundance ([Fe/H]=0,
[E/Fe]=+0.301). From the values given in Table 
\ref{abs} it can be seen that $\alpha$-elements constitute 70\% of
metals at solar abundance ratio. Doubling all these elements
therefore gives $\Delta$[Z/H]=+0.25 (assuming $\Delta$Y=2.2$\Delta$Z; Lebreton et al. 
1999 and an internally consistent handling of H). This yields an estimate 
of $A=0.83$. Non-linear effects and uncertainties in the handling of He result 
in an uncertainty in the value of $A$ of approximately $\pm$ 0.06. 
For solar neighbourhood stars we characterise the trend in [E/Fe] with [Fe/H] in the range
-1$\le$[Fe/H]$<$0 by:

\begin{equation}
[E/Fe]=-0.45[Fe/H].
\label{tea1b}
\end{equation}

\noindent The value 0.45 in Equation \ref{tea1b} is the overall enhancement in
$\alpha$-element abundances with respect to solar
in the data of Salasnich et al \shortcite{Sea00} for low metallicity field 
stars. Combining Equations \ref{tea1a} and \ref{tea1b} we get: 

\begin{equation}
[Z/H]=0.63[Fe/H].
\label{tea1f}
\end{equation}

We use this equation to estimate [Z/H]$_{SSP}$ in SSPs with
-1$\le$[Fe/H]$_{SSP}<$0.

\subsection{Abundance ratio patterns in galaxies}
\label{abratpatt}
Given the well known over-abundance of Mg with respect to Fe in elliptical galaxies
(e.g. O'Connell 1976; Worthey et al. 1992) one approach to modelling abundance
ratios in galaxies would be to use the pattern observed in low-metallicity, 
solar-neighbourhood stars 
which show a similar enhancement in Mg. However, while galaxy studies show  
Mg sensitive indices (Mg$_1$, Mg$_2$ and Mgb) to be enhanced with
respect to Fe indices such as Fe5227, Fe5335 (Gorgas et al 1997;
Vazdekis et al. 1997; K00; Trager et al. 2000a; Fig. \ref{IvsHbeta}
this paper), many of these studies show indices centred on Ca
features (Ca4227, Ca4455) to be un-enhanced (see also Fig.
\ref{CavsHbeta}). 
This is not the only difference between local, low-metallicity stars and galaxy 
populations. C sensitive indices 
(CN$_1$, CN$_2$ and C$_2$4668) are also enhanced in galaxy populations  
(Vazdekis et al. 1997; K00; Fig. \ref{IvsHbeta} this paper). The differences 
between abundance patterns in low-metallicity, solar-neighbourhood stars and 
that in high-metallicity galaxy populations are interesting, but not surprising 
given the difference in both metallicity and environment. They do, 
however, render this approach to modelling galaxy abundance ratios 
unworkable, as many Lick indices are particularly sensitive to C abundance. 
Here we make the assumption that enhancement in Mg abundance reflects an equal 
enhancement in the abundances of all $\alpha$-elements in Table
\ref{abs}, with the
exception of Ca which is assumed to follow Fe peak element abundances. C 
is also assumed to be enhanced. For galaxy population abundances we have 
therefore defined two groups; the 'Fe-like' elements (Ca, Cr, Fe and Ni) 
and the 'enhanced' elements (C, N, O, Ne, Na, Mg, Al, Si, S and Ar -
see Table \ref{abs}, column 4) with 
the abundance of each element, in each group, enhanced by the same factor. This is
similar to model 4 of T00a, where
a similar analysis was carried out (see Section \ref{ESSP}). The two groups of elements
combined represent 99.86\% of the mass of metals present in the solar 
photosphere (Table \ref{abs}). Using these assumptions we hope to obtain 
reasonable estimates of age, metallicity and the degree of enhancement in 
galaxy populations.

For the 3 galaxies with estimated [Fe/H]$<$0 it is necessary to compensate 
for the non-solar abundance ratios in the stars used to calibrate the
SSPs. We use Equation \ref{tea1b} to calculate the
stellar enhancement. This is added to the \emph{estimated} enhancement
(relative to SSP). Finally, Equation \ref{tea1a} and the new value of
[E/Fe] are used to obtain [Z/H].

\subsection{Non-solar abundance ratio isochrones}
\label{nsars}
Until recently, understanding of the effects of non-solar abundance
ratios on stellar populations was poor. However, desire for a better 
description of the evolutionary tracks of globular clusters led to 
the development of theoretical isochrones for $\alpha$-element enhanced 
populations. Salaris 
\& Weiss (1998) and VandenBerg et al. (2000) modelled isochrones for 
the abundance pattern in metal-poor field stars which possess 
enhancements in $\alpha$-element abundances as detailed in Section \ref{lap}. They show 
that, at low metallicities ([Z/H]$<$0), $\alpha$-enhanced isochrones are well estimated by solar 
abundance ratio isochrones of the same [Z/H]. However, they also 
show that at the highest metallicities in their studies ([Z/H]$\sim$solar) this no 
longer holds. Salasnich et al. (2000) modelled $\alpha$-enhanced
isochrones at metallicities similar to those observed in galaxy 
centres ([Z/H]$\sim 0-0.5$). They show that, at these metallicities, due to low 
atomic numbers 
and high ionisation potentials, $\alpha$-elements make relatively small 
contributions to opacity. Consequently,
high metallicity, $\alpha$-enhanced isochrones possess a higher T$_{eff}$ 
than solar abundance ratio isochrones of the same [Z/H]. The resultant 
isochrones are therefore best approximated by solar abundance 
ratio isochrones of lower [Fe/H]. Fig. 10 of Salasnich et al. 
\shortcite{Sea00} shows that an isochrone for a 10 Gyr old population, 
with solar metallicity and an $\alpha$-enhancement of +0.45 (i.e.
[Fe/H]$=-0.37$, 
Equation \ref{tea1a}), is best modelled by a solar abundance ratio isochrone of 
metallicity [Fe/H]$=-0.25$. At [Z/H]=+0.6 ([Fe/H]=+0.23) the effect becomes even more pronounced 
such that the $\alpha$-enhanced isochrone straddles the solar abundance ratio 
isochrones of [Fe/H]=+0.0 and [Fe/H]=+0.3. Consequently, an approximate use 
of the Salasnich et al. \shortcite{Sea00} isochrones is to assume that the 
addition of $\alpha$-elements to high metallicity populations results 
in \emph{no} change in the isochrones. Equivalently, the position of
the isochrones may be considered to depend upon [Fe/H] in the population rather than
[Z/H]. This is clearly only an approximation. However, the 
Salasnich et al. \shortcite{Sea00} isochrones suggest that, at high
metallicities,  it is better than assuming $\alpha$-enhanced
isochrones to be best modelled by solar abundance ratio with the same [Z/H]. 
Differences in abundance pattern between metal-poor field stars and 
galaxy populations are not expected to make a significant difference to the 
behavior of the isochrones as the majority
of significant contributors to opacity are amongst the
elements assumed to show similar abundance ratios in both populations (O, Ne, Na, 
Mg, Al, Si, S, Ar and Ni), while no other individual elements, including
C and Ca, contribute significantly to opacity. 

As SSP indices are 
based on empirical relations for line-strengths versus T$_{eff}$, log g and [Fe/H] 
in stars, the results of Salasnich et al. \shortcite{Sea00} have a
significant impact on the interpretation of SSPs. If it is assumed
that isochrone positions (e.g. T$_{eff}$, log g) are dependent on [Z/H],
then SSPs with [Fe/H]$_{SSP}$=[Z/H]$_{POP}$ (where [Z/H]$_{POP}$ is [Z/H] 
in the galaxy) provide the best match to
enhanced populations. If, on the other hand, it is assumed that 
isochrone position is dependent on [Fe/H], then 
[Fe/H]$_{SSP}$=[Fe/H]$_{POP}$ provides the best match.

\subsection{Estimating indices in non-solar abundance ratio stars}
\label{TRIP}
TB95 modelled the effects of individually doubling the abundances of 10 key 
elements in the synthetic spectra of 3 key stellar types (cool giant, turn-off 
star and cool dwarf), as well as doubling all elements simultaneously.
All models were evaluated at fixed T$_{eff}$ and log g, with values based on a 
5 Gyr, solar metallicity isochrone. TB95 assumed no change in the opacity 
distribution function when doubling individual elements. This approximation 
is appropriate as individual elements contribute little to the opacity. For 
the doubling of all elements, TB95 again assumed fixed T$_{eff}$ and
log g, while an opacity distribution function appropriate for twice
solar metallicity was used. The results are 
presented as variations of the 21 Lick indices modelled, in terms of a standard 
error, caused by doubling each of the 10 elements. The results of doubling 
all elements are presented in the same manner. Indices modelled by
TB95 include 16 present in our study. TB95 did not, however, model H$\delta$, H$\gamma$ 
MgI or CaT. One of the elements modelled by TB95 was Ti. The abundance
of this element is enhanced in low
metallicity stars while it's atomic mass lies in the range of the Fe-like elements. 
Consequently, the decision as to whether to include Ti among the enhanced or
Fe-like indices is difficult. We have therefore not included this 
(low abundance) element in our analysis. TB95 is the only published
study of this type to date.

The effects of changing abundance ratios can be estimated from the
TB95 data by defining $R_{i,X}$; 
the fractional change in the ith index in the TB95 arrays (e,g, CN$_1$, CN$_2$, 
Ca4227 etc) when the abundance of the element X (e.g. C/H, N/H) is doubled. 
The value of the enhanced index ($I_{i}^{'}$) can then be estimated
from the solar abundance ratio value (I$_i$) by (following T00a):

\begin{equation}
I_{i}^{'}=I_i[(1+R_{i,X_1})^{E_{X_1}/log 2}(1+R_{i,X_2})^{E_{X_2}/log 2}........]
\label{combine}
\end{equation}

Where E$_X$ is the change in abundance of element X, i.e.
E$_X$=$\Delta$[X/H]. 
Doubling all elements (column 14 of TB95; 
Tables 4 to 6) can be handled in the same way i.e. E$_Z$=$\Delta$[Z/H].
It should be noted that the doubling of all elements in TB95 did not
model a simple doubling of the stellar metallicity as the TB95 calculations
were made assuming \emph{fixed} photospheric conditions (T$_{eff}$, log g).

\subsection{Constructing non-solar abundance ratio SSPs}
\label{ESSP}
\begin{table*}
\begin{center}
\caption{Fractional changes (R$_{i,X}$) in index values of 53/44/3 mix when
abundances of individual elements (e.g. C/H) are doubled. These data are
derived from Tables 4 to 6 of TB95 and are shown here to illustrate
the sensitivities to individual elements.}
\begin{tabular}{lrrrrrrrrrrrrr}
\hline
Index  &C&N&O&Mg&Fe&Ca&Na&Si&Cr&Ti&Z\\
\hline
CN$_1$ &  1.75&  0.54& -0.46& -0.13& -0.03& -0.07& -0.03&  0.11& -0.10&  0.03&  0.40\\
CN$_2$ &  1.09&  0.35& -0.29& -0.10& -0.03& -0.05& -0.02&  0.12& -0.05&  0.03&  0.29\\
Ca4227 & -0.32& -0.05&  0.10&  0.00&  0.05&  0.31& -0.01&  0.00& -0.01&  0.00&  0.24\\
G4300  &  0.27&  0.00& -0.06& -0.02& -0.04&  0.01& -0.01& -0.01& -0.02&  0.05&  0.04\\
Fe4383 &  0.08& -0.00& -0.02& -0.05&  0.20& -0.03& -0.01& -0.04&  0.00&  0.02&  0.12\\
Ca4455 & -0.05& -0.01&  0.01& -0.01& -0.08&  0.00& -0.01& -0.00&  0.07&  0.03&  0.16\\
Fe4531 &  0.00&  0.01&  0.01& -0.01&  0.03&  0.00&  0.01& -0.05&  0.04&  0.11&  0.14\\
C$_2$4668&1.90&  0.00& -0.27& -0.06&  0.03& -0.01& -0.01& -0.12& -0.02&  0.04&  0.36\\
H$\beta$& 0.03&  0.00& -0.00& -0.04&  0.00&  0.00&  0.01&  0.01& -0.04&  0.00& -0.00\\
Fe5015 & -0.00&  0.00& -0.00& -0.10&  0.09&  0.01&  0.01& -0.03& -0.02&  0.08&  0.14\\
Mg$_1$ &  0.78& -0.00& -0.11&  0.26& -0.10& -0.01& -0.01& -0.05& -0.01&  0.02&  0.21\\
Mg$_2$ &  0.11& -0.01& -0.03&  0.23& -0.04& -0.01& -0.01& -0.03& -0.01&  0.02&  0.14\\
Mgb    & -0.17& -0.01& -0.00&  0.37& -0.07& -0.00& -0.01& -0.04& -0.10& -0.01&  0.08\\
Fe5270 &  0.07&  0.01& -0.00& -0.05&  0.11&  0.02& -0.01& -0.01&  0.01&  0.02&  0.14\\
Fe5335 & -0.05& -0.01&  0.01& -0.04&  0.20&  0.00& -0.01& -0.00&  0.03&  0.02&  0.14\\
Fe5406 &  0.03&  0.01& -0.00& -0.01&  0.17& -0.01& -0.01& -0.00&  0.05&  0.02&  0.15\\
\hline
\label{fracts}
\end{tabular}
\end{center}
\end{table*}

It is possible to estimate the effects of non-solar abundance ratios
in populations by modelling SSPs as combinations of the cool 
giant/turn-off/cool dwarf stellar types of TB95. The luminosity-weighted 
sum of the factors in Tables 4 to 6 of TB95 allows estimation 
of R$_{i,X}$ values for SSPs. These can then be
combined, by use of Equation \ref{combine}, to estimate the
sensitivities of indices in SSPs to enhancement of elements in the 
chosen abundance ratio pattern. Trager et al. 
\shortcite{Tea00a,Tea00b} used a 53/44/3 percentage luminosity-weighted
combination of the three stellar types to simulate SSPs. Although not detailed in 
their paper, the 53/44/3 combination is consistent with values given
by W94 for the relative contributions of these stellar types
to 3-17 Gyr SSPs of [Fe/H]$_{SSP}\sim$0. The similarity in the sensitivities (in percentage 
terms) of cool giant and turn-off stars make the results fairly robust to any 
reasonable combination. We have therefore adopted the T00a
luminosity weightings throughout this paper for ease of comparison. Fractional 
changes in indices for doubling individual element abundances, and doubling
[Z/H], for this combination of stellar types, are given in Table 
\ref{fracts}. The sensitivities of indices to individual elements
divides them into 2 groups; those with sensitivities to
individual elements significantly larger than their sensitivity to Z 
(CN$_1$, CN$_2$, G4300, C$_2$4668, Mg$_1$ and Mgb) and
those whose sensitivities to individual elements are similar to or
less than their
sensitivity to Z (the rest). It is interesting to note that 
the strong sensitivities of the first group are mainly to C and Mg, both of 
which are enhanced elements in galaxy populations (e.g. see Fig.
\ref{IvsHbeta}, right side). Consequently, in
galaxies, these indices are dominated by abundance ratio effects rather 
than [Z/H]. 

The next step in constructing non-solar abundance ratio SSPs is to select the SSP 
whose isochrone best matches that expected in the population under
study. It is to the indices of this SSP that the TB95 corrections for
non-solar abundance ratios are applied. On the basis of the isochrone
models available at the time \cite{SW98,Vea00}, T00a assumed that the isochrone 
position was governed by [Z/H], thus an SSP with 
[Fe/H]$_{SSP}$=[Z/H]$_{POP}$
was selected. However,  if we accept the implications of the subsequent Salasnich et al. 
\shortcite{Sea00} study - that isochrone positions are dependent on [Fe/H] rather than
[Z/H] - then we must select an SSP with [Fe/H]$_{SSP}$=[Fe/H]$_{POP}$. Given the
uncertainties in this aspect of the modelling we tested three methods 
for applying the TB95 data to galaxy observations; the method used by
T00a, which assumes isochrone shape to be governed by [Z/H], and two methods designed to be
consistent with Salasnich et al. \shortcite{Sea00} isochrones.

\subsubsection{T00a method}
Given the assumption that isochrone positions are governed by [Z/H], T00a 
pointed out that there is only one way to achieve the 
required element abundance ratio enhancements. This involves the reduction in the abundance 
of Fe-like elements, while enhanced element abundances are (marginally) 
increased to maintain [Z/H]. Abundances of elements not modelled 
by TB95 are assumed to remain constant. T00a used Equation \ref{tea1d}
to derive:

\begin{equation}
\Delta[Fe/H]=-A\Delta[E/Fe], \hspace{3mm} \Delta[Z/H]=0
\label{tea1c}
\end{equation}

Where E now refers to all elements enhanced with respect to Fe-like
elements in galaxy populations. The data in Table \ref{abs}, column 4
leads to a value of $A=0.942$ for the galaxy abundance ratio pattern using the 
T00a method. From Equation \ref{tea1c} this method requires a value 
of E$_X$ (in Equation \ref{combine}) given by:

\begin{equation}
E_X=(1-A)\Delta[E/Fe], \hspace{3mm} X=C,N,O,Mg,Na,Si
\label{tea1e}
\end{equation}

\noindent While for Fe-like indices:

\begin{equation}
E_X=-A\Delta[E/Fe], \hspace{3mm} X=Fe,Ca,Cr
\label{tea1g}
\end{equation}

As this method directly estimates age, [Z/H] and [E/Fe], Equation
\ref{tea1a} is used to calculate [Fe/H]. The difficulty with this method 
is that it is based on the (now apparently false) assumption that,  
at the high metallicities observed in galaxy centres, isochrone
positions are governed by [Z/H]. Nonetheless, we have applied the T00a analysis 
to our data using V99 SSPs.

\subsubsection{Methods based on Salasnich isochrones}
In light of the Salasnich et al. \shortcite{Sea00} isochrones we chose 
to test two alternative approaches to the application of TB95 data. As these 
methods are based on the assumption that, at the high metallicities
present in the majority of galaxies, isochrone shape is governed by [Fe/H], 
we select the SSP with [Fe/H]$_{SSP}$=[Fe/H]$_{POP}$. 
We then ensure that the applied enhancements involve no 
change in the abundance of Fe-like elements. We identify two ways of achieving these 
requirements with the TB95 data. The first is simply to increase the abundance of 
each of the elements thought to be enhanced in galaxy populations, while keeping 
other element abundances constant; the E+ method. For this method, the values of 
E$_X$, used in Equation \ref{combine}, are given by:

\begin{equation}
E_X=\Delta[E/Fe], \hspace{3mm} X=C,N,O,Mg,Na,Si
\label{tea1h}
\end{equation}

\noindent For other elements and Z in Table \ref{fracts} E$_X$=0. As this method is
specifically designed to reflect high metallicity isochrones, it is
not used  for comparisons with observations of three low metallicity 
([Z/H]$<$0) bulges. Instead, for these galaxies, we use the T00a method which
is consistent with the low metallicity isochrones of  Salaris \& Weiss 
(1998) and VandenBerg et al. (2000). Metallicities and abundance ratios in these
three galaxies are transformed to solar scale as described in Section
\ref{abratpatt}.

The second approach is to double \emph{all} elements (Z; Table \ref{fracts}) 
then \emph{reduce} the abundances of Fe-like elements (Fe, 
Ca and Cr); the Fe$-$ method. This method effectively doubles all elements 
except those in the Fe-like group. 
It should again be noted that doubling Z in the tables of TB95
does not represent a simple doubling of the metallicity
in a real population, as the TB95 calculations were carried out at fixed
T$_{eff}$ and log g, i.e. with no movement of the isochrone. However, 
this is exactly the requirement of this method, as it assumes that 
addition of elements whose abundances are enhanced in galaxy populations leaves isochrone 
positions unchanged. For the Fe$-$ method, the values of E$_X$ used in Equation
\ref{combine} are given by:

\begin{equation}
E_X=\Delta[E/Fe], \hspace{3mm} X=Z
\label{tea1i}
\end{equation}

\noindent and 

\begin{equation}
E_X=-\Delta[E/Fe], \hspace{3mm} X=Fe,Ca,Cr
\label{tea1j}
\end{equation}

\noindent Again we have used the T00a method for three bulges with [Z/H]$<$0. 
Despite the differences between T00a and Fe$-$ methods, there 
are strong similarities, as the fractional changes 
calculated for both methods are dominated by the reduction in Fe-like
elements. However, the assumptions regarding isochrones do result in
significant differences between estimated ages, as we shall see in
Section \ref{rescomps}.\\

As both E+ and Fe$-$ methods are based on SSPs of known [Fe/H], rather than [Z/H], we must
re-write Equation \ref{tea1d} as:

\begin{equation}
\Delta[Z/H]=A\Delta[E/Fe], \hspace{3mm} \Delta[Fe/H]=0
\label{tea1k}
\end{equation}

The data in Table \ref{abs}, column 4 yields $A=0.941$ for both E+
and Fe$-$ methods. As both these methods directly estimate age, [Fe/H] 
and [E/Fe], Equation \ref{tea1a} is used to calculate [Z/H].

\subsubsection{Comparison of results from different methods}
\label{rescomps}
For all three methods, grids of non-solar abundance ratio SSPs were generated for [E/Fe] 
ranging from $-$0.3 to +0.6 in steps of 0.025 dex. Grid spacings of
0.0125 dex were used for 0.175$\leq$log(age)$\leq$1.225 and 0.025 dex for 
$-0.5\leq$[Fe/H]$\leq0.75$. Linear extrapolation of the V99 data
from [Fe/H]=+0.4 to +0.75 was necessary due to the high Fe index
values of a single  S0 galaxy (NGC 2549).
Indices not modelled by TB95 (H$\delta_A$, H$\delta_F$, 
H$\gamma_A$ and H$\gamma_F$, MgI and CaT) were assumed to have no
sensitivity to abundance ratio. The 
best fit (minimum $\chi^2$) was found for each galaxy, for each method. For the 6 
galaxies observed to have aberrant emission (Section \ref{ecorr}), H$\beta$ 
was omitted from the minimisation procedure. The MgI and CaT indices were also omitted. 

\begin{table}
\caption{Comparison of Fe$-$, E+ and T00a methods for modelling
non-solar abundance ratio SSPs. Mean deviations from the best fit
enhanced SSP models are given  (see text for details).}
\begin{tabular}{lrrrrrrr}
\hline
Index        & N &\multicolumn{2}{c}{{Fe$-$}}&\multicolumn{2}{c}{{E+}}
&\multicolumn{2}{c}{{T00a}}\\  
         &      & Mean  &  $\chi^2$ & Mean &   $\chi^2$& Mean &  $\chi^2$\\
\hline
H$\delta_{A}$& 32&  1.1&     74&  1.9&    160&  2.1 &   179 \\
H$\delta_{F}$& 32&  1.0&     81&  2.2&    228&  2.1 &   212 \\
CN$_{1}$     & 32&  0.5&     32&  0.4&     25&  0.1 &    21 \\
CN$_{2}$     & 32&  0.7&     53&  0.5&     38&  0.4 &    38 \\
Ca4227       & 32& -2.8&    378& -3.3&    485& -1.7 &   194 \\
G4300        & 32&  0.2&     52&  0.1&     49&  0.1 &    51 \\
H$\gamma_{A}$& 32&  0.6&     49&  1.1&     79&  1.0 &    80 \\
H$\gamma_{F}$& 32&  0.3&     22&  0.6&     32&  0.5 &    29 \\
Fe4383       & 32&  0.6&     26&  0.4&     19&  0.5 &    21 \\
Ca4455       & 32& -1.0&     50& -0.6&     22& -1.0 &    44 \\
Fe4531       & 32&  0.2&     16&  0.3&     18&  0.4 &    20 \\
C$_{2}$4668  & 32& -0.3&     50& -0.9&     74& -0.1 &    39 \\
H$\beta$     & 26& -1.4&    141&  0.5&    128& -1.2 &   139 \\
Fe5015       & 32& -0.6&    139&  1.2&    127& -0.1 &   109 \\
Mg$_{1}$     & 32& -0.1&    102& -2.1&    212& -0.1 &    97 \\
Mg$_{2}$     & 32&  1.2&    128&  1.8&    187&  1.4 &   152 \\
Mgb          & 32& -0.7&     89&  6.2&   1792& -0.8 &    98 \\
Fe5270       & 32&  0.4&     67& -1.4&    198&  0.7 &    72 \\
Fe5335       & 32&  0.2&     55& -1.6&    163& -0.5 &    70 \\
Fe5406       & 30&  1.9&    165& -0.3&     24&  1.3 &    91 \\
CaT$^\dag$   & 26& -3.2&  (507)& -2.8&  (376)& -5.5 &(1189) \\
MgI$^\dag$   & 26&  6.8& (1374)&  6.5& (1287)&  6.4 &(1229) \\
\hline
Total        &   &     &   1769&     &   4060&      &  1756 \\
\hline	
\label{methcomps}
\end{tabular}
$^\dag$ Not included in $\chi^2$ minimisations.
\end{table}

\begin{figure}
\vspace{3cm}
Figure available from \\
http://www.star.uclan.ac.uk/\~{}rnp/research.htm
\vspace{4cm}
\caption{Comparison of results obtained using the T00a method to 
those using the Fe$-$ method (see Section 4.5). Open circles are
elliptical galaxies, open squares are S0s and filled symbols are 
late-type galaxies.}
\label{cfteamy}
\end{figure}

Table \ref{methcomps} shows the average deviation 
(difference between observed and best fit non-solar abundance ratio
SSP indices, as a multiple of our
observational error) and $\chi^2$ values for each index, over all 
observed galaxies, using each of the three methods. Comparison of best fits 
for E+ and Fe$-$ methods shows the Fe$-$ method to have a significantly 
lower total $\chi^2$. We therefore dismiss the E+ method as too unrealistic.
The T00a method has marginally lower total $\chi^2$ than the Fe$-$ method. 
However, this difference hinges on a single index (Ca4227) whose error is 
underestimated (Section \ref{flux}). If this index is excluded the Fe$-$ 
method possesses a total $\chi^2$ value $\sim$ 12\% lower than the
T00a method. Comparison of the results for estimates of luminosity-weighted 
log(Age), [Fe/H], [Z/H] and [E/Fe] from T00a and Fe$-$ methods are 
shown in Fig. \ref{cfteamy}. It can be seen that for [Fe/H], [Z/H] and
[E/Fe] the results for the two methods are in good agreement. However, 
the T00a method  gives values of log(Age) significantly lower (by $\leq$
0.25 dex) than those derived
by the Fe$-$ method. This is a direct result of the difference in assumptions 
about the effects of non-solar abundance ratios on isochrone positions 
detailed in Section \ref{ESSP}. The log(Age) ordering agrees fairly well, to
within $\sim$ 0.1 dex. Table \ref{agez} presents results derived by the Fe$-$ method for all galaxies with
[Z/H]$>$0, to be consistent with the recent Salasnich et al.
\shortcite{Sea00} isochrones. For galaxies with [Z/H]$<$0 the T00a
method has been applied in line with the low metallicity isochrones 
of Salaris \& Weiss (1998) and VandenBerg et al. (2000).

\begin{table*}
\caption{Central values of luminosity-weighted log(Age), [Fe/H], [E/Fe] and
[Z/H] for the 32 galaxies in our sample. Extent of 1 $\sigma$
confidence contours are given as errors (in brackets). Data points with
an asterisk are omitted from Fig. 12.}
\begin{tabular}{ccccc}
\hline
   Galaxy & Log(Age)& [Fe/H ]& [E/Fe]  &   [Z/H]   \\
\hline
Ellipticals&&&&\\
\hline   
 NGC2832 &   0.875(0.069)&  0.175(0.050)&  0.300(0.025)&  0.457(0.074)\\
 NGC2831 &   0.613(0.069)&  0.075(0.050)&  0.250(0.038)&  0.310(0.085)\\
 NGC3226 &   1.025(0.056)& -0.075(0.038)&  0.425(0.025)&  0.325(0.061)\\
 NGC3608 &   0.950(0.044)&  0.125(0.025)&  0.275(0.013)&  0.384(0.037)\\
 NGC4291 &   1.013(0.050)&  0.000(0.038)&  0.300(0.025)&  0.282(0.061)\\
 NGC4365 &   0.988(0.044)&  0.175(0.025)&  0.250(0.013)&  0.410(0.037)\\
 NGC4374 &   1.138(0.056)&  0.000(0.038)&  0.300(0.025)&  0.282(0.061)\\
 NGC4552 &   0.988(0.044)&  0.150(0.038)&  0.325(0.025)&  0.456(0.061)\\
 NGC4636 &   0.913(0.063)&  0.175(0.038)&  0.275(0.025)&  0.434(0.061)\\
 NGC4697 &   0.712(0.075)&  0.300(0.025)&  0.200(0.025)&  0.488(0.049)\\
 NGC5322 &   0.625(0.056)&  0.250(0.025)&  0.150(0.025)&  0.391(0.049)\\
\hline
S0s&&&&\\
\hline   
 NGC2549 &   0.313(0.113)&  0.600(0.075)&  0.150(0.013)&  0.741(0.087)\\
 NGC3607 &   0.750(0.056)&  0.250(0.025)&  0.250(0.025)&  0.485(0.049)\\
 NGC4203 &   0.813(0.063)&  0.225(0.038)&  0.300(0.025)&  0.507(0.061)\\
 NGC4526 &   0.587(0.038)&  0.325(0.038)&  0.175(0.025)&  0.490(0.061)\\
 NGC5354 &   0.688(0.113)&  0.225(0.062)&  0.250(0.038)&  0.460(0.098)\\
 NGC5353 &   0.863(0.081)&  0.300(0.050)&  0.275(0.025)&  0.559(0.074)\\
\hline
Bulges&&&&\\
\hline    
 NGC2683 &   0.675(0.088)& -0.025(0.050)&  0.200(0.038)&  0.163(0.085)\\
 NGC3254 &   0.587(0.056)&  0.025(0.050)&  0.150(0.025)&  0.166(0.074)\\
 NGC3301 &   0.338(0.050)&  0.175(0.050)&  0.075(0.013)&  0.246(0.062)\\
 NGC3623 &   0.725(0.069)&  0.300(0.037)&  0.125(0.013)&  0.418(0.049)\\
 NGC3769 &   $<$0.175(0.013)$^*$&-0.576(0.013)$^*$&0.234(0.113)$^*$&-0.356(0.346)$^*$\\
 NGC4157 &   0.512(0.169)& -0.168(0.138)&  0.201(0.087)&  0.021(0.220)\\
 NGC4192 &   0.300(0.056)&  0.350(0.038)&  0.150(0.025)&  0.491(0.061)\\
 NGC4216 &   0.600(0.025)&  0.400(0.025)&  0.200(0.025)&  0.588(0.049)\\
 NGC4217 &   0.463(0.363)$^*$&-0.050(0.438)$^*$&0.022(0.275)$^*$&-0.029(0.697)$^*$\\
 NGC4312 &   0.463(0.213)$^*$&-0.449(0.213)$^*$&0.227(0.237)$^*$&-0.235(0.436)$^*$\\
 NGC4313 &   $<$0.175(0.050)$^*$& 0.325(0.050)$^*$&0.075(0.038)$^*$& 0.396(0.085)$^*$\\
 NGC4419 &   0.313(0.044)&  0.050(0.050)&  0.175(0.038)&  0.215(0.085)\\
 NGC5746 &   0.750(0.125)&  0.250(0.050)&  0.175(0.025)&  0.415(0.074)\\
 NGC5908 &   0.475(0.069)&  0.350(0.075)&  0.125(0.025)&  0.468(0.099)\\
 NGC5987 &   0.500(0.056)&  0.350(0.075)&  0.150(0.025)&  0.491(0.099)\\
\hline
\label{agez}
\end{tabular}
\end{table*}

\begin{table*}
\caption{Comparison of results of age/metallicity estimates omitting
differing groupings of indices. Average offset and RMS scatter about offset (in brackets) are given for each of the
derived parameters. Also shown (where appropriate) are the scatters about the most significant correlations detected (see
Tables \ref{Acorrels} and \ref{Scorrels}). N.B. 6 galaxies with
aberrant emission (Section 2.5.2) were excluded from this analysis. }
\begin{tabular}{lcrrrrccc}
\hline                                                                     
         &    &                &              &              &             & Early Type&  Early Type&   Late Type\\
Indices  &    &                &              &              &             & log(Age)  &  log(Age)  &   log($\sigma$) \\
omitted  & N  &$\Delta$log(Age)&$\Delta$[Fe/H]&$\Delta$[E/Fe]&$\Delta$[Z/H]& vs [Fe/H] & vs [E/Fe]  &   vs [Fe/H]\\
\hline
\multicolumn{9}{l}{{\bf WITH ENHANCEMENT}}\\
None                            & 20 &  0.000(0.000)  & 0.000(0.000) & 0.000(0.000) & 0.000(0.000)&    0.089  &     0.041  &     0.087\\
H$\delta_F$,H$\gamma_F$         & 18 & -0.020(0.026)  & 0.005(0.016) & 0.008(0.021) & 0.012(0.020)&    0.089  &     0.037  &     0.077\\
H$\delta_{A,F}$,H$\gamma_{A,F}$ & 16 & -0.030(0.044)  & 0.002(0.023) & 0.013(0.020) & 0.015(0.022)&    0.084  &     0.031  &     0.073\\
H$\beta$                        & 19 &  0.021(0.066)  &-0.011(0.037) &-0.004(0.026) &-0.015(0.050)&    0.087  &     0.050  &     0.098\\
H$\beta$,H$\delta_{A,F}$,H$\gamma_{A,F}$&15&-0.007(0.090)&-0.009(0.042)&0.006(0.027)&-0.003(0.056)&    0.091  &     0.059  &     0.102\\ 
All except                      & 3& -0.070(0.171)  & 0.004(0.129) & 0.038(0.047) & 0.040(0.156)&    0.094  &     0.055  &     0.098\\
H$\beta$, Mg$_2$,$<$Fe$>$       &    &                &              &              &             &           &            &          \\

\hline	 
\multicolumn{9}{l}{{\bf WITHOUT ENHANCEMENT}}\\
None  & 20 & -0.022(0.057)  &     -        &     -        &0.035(0.028) &     -     &      -     &       -    \\    
All except& 2&  0.041(0.157)  &     -        &     -        &0.039(0.128) &     -     &      -     &       -\\
H$\beta$, [MgFe]&    &                &              &              &             &           &            &          \\

\hline                                                                     
\label{enablecomps}
\end{tabular}
\end{table*}

As a check on our decision to include all Balmer line indices (H$\beta$, H$\gamma_{A,F}$ and
H$\delta_{A,F}$) in the derivation of the values given in Table
\ref{agez}, estimates were also made with differing combinations
of these 5 indices excluded from the fitting procedure. Table
\ref{enablecomps} details the comparisons of these age/metallicity estimates with 
those given in Table \ref{agez}. Also shown are the scatters of the
derived values about key correlations identified in Section \ref{results}.
Offset and scatter in the derived
values, when combinations of Balmer lines are excluded from the fitting
procedure, are small, with  values $<$0.05 dex in cases when
H$\beta$ is not omitted. Indeed, even when all Balmer line indices
are omitted, agreement with the values in Table \ref{agez} is
relatively good.
The comparisons presented in Table \ref{enablecomps}, therefore, show that 
our results are robust to the choice of Balmer lines to include 
in the fitting procedure. The results even suggest that, with a 
large number of indices, reasonable estimates can be made
\emph{without} the Balmer lines. As many studies have been based on
combinations of indices such as H$\beta$, Mg$_2$ and $<$Fe$>$, we have
also compared results
from this combination in Table \ref{enablecomps}. We find only modest
offsets. However, the scatter is large ($\sim$ 0.15 dex). Scatter
about the key correlations are also found to be increased in this case.
These results emphasise the
advantage of the large number of indices included in this study.

The effects on the results of ignoring enhancements are also shown in
Table \ref{enablecomps}. 
To obtain this comparison, the best fits of our data (all 20 indices) to
un-enhanced SSP index values were found. It was assumed
that [Fe/H]$_{SSP}$ of the best fit SSP best represents the value of [Z/H]
for the population. 
Comparison of the values of log(Age) and [Z/H], estimated in this
way, to the values given in Table \ref{agez} (derived by the Fe$-$
method), show that the  average difference between values are $<$0.05
dex in both parameters. Scatter about these differences are
similarly $\lesssim$0.05 dex. We also tested the method employed
by some authors (e.g. Gonzalez 1993; Kuntschner \& Davies 1998) of combining the Mgb
and $<$Fe$>$ indices by taking their geometric mean ($\sqrt{Mgb<Fe>}$)
to make a new index ([MgFe]). Age and metallicity estimates are then
made using this index and an age sensitive index; normally H$\beta$.
Results of the comparison between log(Age) and [Z/H] values derived using
this method, and those derived from the Fe$-$ method, are also shown in 
Table \ref{enablecomps}. The average differences between the two sets
of values is similar in magnitude to those found above for all 20
indices (without enhancement). However, the scatter in the differences is 
significantly greater ($\sim$ 0.15 dex). This result again emphasises
the benefit of including a large number of indices in the fitting
procedure.

\subsection{Results from blue indices}  
\label{bresults}
\begin{figure*}
\vspace{3cm}
Figure available from http://www.star.uclan.ac.uk/\~{}rnp/research.htm
\vspace{20cm}
\caption{Luminosity-weighted [Fe/H], [E/Fe] and [Z/H] are plotted
against log(Age). Recall E represents the enhanced element group (Table 7, column 4).
Data identified with an asterisk in Table 10 are omitted from
this plot. Correlations for early-type galaxies 
are shown as lines. Key: $\bullet$=bulges, $\Box$=S0, $\bigcirc$=E, 
$\times$=Virgo cluster galaxies. 1 $\sigma$ confidence contours from our fits
are shown on the right.}
\label{Age}
\end{figure*}


\begin{figure}
\vspace{3cm}
Figure available from \\
http://www.star.uclan.ac.uk/\~{}rnp/research.htm
\vspace{19cm}
\caption{Luminosity-weighted log(Age), [Fe/H], [E/Fe] and [Z/H] are plotted
against the log of central velocity dispersion. Symbols as Fig. 12.}
\label{Sig}
\end{figure}

Values for luminosity-weighted log(Age), [Fe/H], [Z/H] and [E/Fe] derived by the Fe$-$ 
method detailed above, are given in Table \ref{agez}. Plots of
[Fe/H], [Z/H] and [E/Fe] against log(Age) are shown in Fig. \ref{Age}. 
To avoid confusion points with errors $>$0.2 dex, or limits, (see Table \ref{agez}) have 
been omitted from these plots.
Confidence contours (1 $\sigma$, allowing for 3
interesting parameters) are plotted on the right of Fig. \ref{Age} as quadrilaterals 
with vertices at extremes projected onto the plane presented. The 
alignment of the contours in plots of [Z/H] and [Fe/H] against log(Age) clearly show 
the age/metallicity degeneracy. However, the size of the contours suggests 
that, for the majority of galaxies, the degeneracy has been broken.

It can be seen in Fig. \ref{Age} that the ages of spiral bulges, S0s and Es form a 
continuous, overlapping sequence of increasing luminosity-weighted age, with 
bulges ranging from 1.5 to 6 Gyr old, while S0s and Es range from 2 to 7 
Gyr and 4 to 13 Gyr respectively. S0s also appear more Fe rich than
Es. This is reflected in the strong central Fe features seen in S0s (e.g.
Figs \ref{Iwithsig} and \ref{Htype}).

We fitted lines (by $\chi^2$ minimisation) to both the early and late-type data plotted
in Fig. \ref{Age}. The four galaxies identified by asterisks in Table
\ref{agez} were omitted from the fitting procedure.
Table \ref{Acorrels} give the results of our fits.
These have been calculated as the average of the fits obtained when
y-axis  and x-axis deviations are minimised separately. Errors
quoted in Table \ref{Acorrels} are half the difference between the
values from these two fits. Significant correlations are shown as
lines in Fig. \ref{Age}. We find a strong anti-correlation between [Fe/H] and
luminosity-weighted log(Age) in early-type galaxies. However, it was a
concern that this may be heavily influenced by the low age, S0 galaxy
(NGC 2549). We therefore re-calculated the correlation
omitting this galaxy. A slope of 0.672$\pm$0.193 was found indicating 
that this point is not having an excessive influence (cf Table \ref{Acorrels}). 
We also detect a strong correlation between [E/Fe] and log(Age) in early-type galaxies. 
These trends are reflected in the index-index plots of Fig. \ref{IvsHbeta} (Section 
\ref{break}). The anti-correlation between [Z/H] (calculated by Equation \ref{tea1a}) 
and log(Age) is a natural consequence of the somewhat stronger correlations with [Fe/H] 
and [E/Fe]. Both the (relatively young) S0s and the 5 Virgo cluster, ellipticals
(which are amongst the oldest in our sample) appear to follow the same trends as the 
early-type sample as a whole. We find that elliptical galaxies
tend to be older and more metal-poor than S0 galaxies.
No significant correlations were found in 
late-type galaxies. However, the [E/Fe]$-$log(Age) plot shows that the data for
early and late types form a continuous monotonic locus. This
correlation may then be common to both Hubble types.

Luminosity-weighted log(Age), [Fe/H], [Z/H] and [E/Fe] are plotted against 
log of central velocity dispersion in Fig. \ref{Sig}. Fits to these
data (minimising $\chi^2$) are
given in Table \ref{Scorrels}. Values given are from fits with y-axis 
errors  only (which dominate in most cases). However, errors 
given are half the difference between the 
quoted fit and x-axis error only fit. Correlations with $\geq 3 \sigma$
significance are shown as lines in Fig. \ref{Sig}. There is a trend 
for luminosity-weighted age to increase with central velocity dispersion 
in our early-type galaxy sample. 
This trend is also consistent with the late-type data. However, we
find no significance in the correlation of age versus velocity dispersion in
late types (Table \ref{Scorrels}). The trend of increasing [E/Fe] with 
velocity dispersion for early-type galaxies suggested by Fig.
\ref{Sig} also has very low significance particularly when compared to
the strength of the correlation of [E/Fe] with log(Age) in early-types. Our data therefore
suggest that, in the early-type galaxies in our sample, the stronger 
correlations of both [E/Fe] and velocity dispersion with log(Age) are the cause of the apparent
trend of [E/Fe] with velocity dispersion. Bulges, on the other hand, 
while showing no significant correlations with age, show strong 
correlations of [Fe/H] and [Z/H] with velocity dispersion and only marginal 
(approximately constant) enhancement with respect to solar, i.e. [E/Fe] 
$\sim$ +0.15. Thus our results suggest that while the principle parameter
controlling abundances and abundance ratios in early-type galaxies is
age, in bulges it is central velocity dispersion.\\

We have compared the trends in our early-type galaxy sample with the
results of Trager et al. (2000b) and Kuntschner et al. (2001). 
Both authors find an anti-correlation between age and metallicity that
is approximately aligned with the age/metallicity degeneracy. 
These correlations are in good agreement with the anti-correlation 
found in our early-type sample (Fig.
\ref{Age}). Trager et al. (2000b) also found an offset of the 
age$-$metallicity relation with lower velocity dispersion. Future
spectral observations of low velocity dispersion ellipticals with high
signal-to-noise are needed to further investigate these trends, incorporating 
large numbers of indices.

\begin{table}
\caption{Correlations with log(Age). The number of galaxies fitted (N)
and best fit slopes and intercepts (errors in brackets) are given.
Also shown are the (unweighted) correlation coefficient (r) and
$\chi^2$ values.}
\begin{tabular}{lrrrrr}
\hline
 & N & Slope & Intercept & r & $\chi^2$ \\
\hline
E/S0s &&&&&\\
\hline   
$[Fe/H]$ &   17 & -0.701(0.167) &    0.771(0.139) & -0.81 & 71\\
$[E/Fe]$ &   17 &  0.346(0.116) &   -0.030(0.099) &  0.78 & 65\\
$[Z/H]$  &   17 & -0.551(0.243) &    0.893(0.202) & -0.66 & 70\\
\hline
Bulges&&&&&\\
\hline
$[Fe/H]$ &   11 &  1.658(1.604) &   -0.582(0.814) & -0.01 &93\\
$[E/Fe]$ &   11 &  0.364(0.253) &   -0.039(0.11)  &  0.34 &45\\
$[Z/H]$  &   11 &  1.538(1.410) &   -0.387(0.702) &  0.06 &55\\
\hline
\label{Acorrels}
\end{tabular}
\end{table}

\begin{table}
\caption{Correlations with log($\sigma_0$). The number of galaxies fitted (N) 
and best fit slopes and intercepts (errors in brackets) are given. 
Also shown are the (unweighted) correlation coefficient (r) and 
$\chi^2$ values.}
\begin{tabular}{lrrrrr}
\hline
 & N &Slope&Intercept&r&$\chi^2$   \\
\hline
E/S0s&&&&&\\
\hline   
Log(Age) &   17 &  1.655(0.690)&    -3.079(1.642) &  0.74 &110\\
$[Fe/H]$ &   17 & -0.727(0.881)&     1.906(2.089) & -0.55 &159\\
$[E/Fe]$ &   17 &  0.422(0.461)&    -0.735(1.091) &  0.39 &117\\
$[Z/H]$  &   17 & -0.553(0.688)&     1.736(1.631) & -0.53 & 31\\
\hline
Bulges&&&&&\\
\hline
Log(Age) &  11 &  0.811(1.053) &   -1.265(2.277) & 0.54 & 38\\
$[Fe/H]$ &  11 &  1.043(0.495) &   -2.023(1.093) & 0.80 & 49\\
$[E/Fe]$ &  11 &  0.246(0.423) &   -0.401(0.917) & 0.04 & 25\\
$[Z/H]$  &  11 &  1.177(0.338) &   -2.169(0.750) & 0.85 & 15\\
\hline
\label{Scorrels}
\end{tabular}
\end{table}

\subsection{Results from red indices}
\label{rresults}
Estimation of age and metallicity from the NIR data is
difficult, as the only age sensitive indices are in the blue wavelength
range. We must therefore combine indices from populations that may
differ, e.g. H$\beta-$CaT in Fig. \ref{CavsHbeta}. The NIR indices
were also not modelled by TB95, so abundance ratio effects can not be
estimated. We have nonetheless derived log(Age) and [Fe/H] estimates from the calcium
triplet indices and H$\beta$. The MgI index was excluded as, contrary
to expectation, the positions of values of this index with respect to
the SSP grids imply \emph{lower} metallicities than do values of the 
CaT index. Many of the early-type galaxy data points also imply
ages greater than 17 Gyr. We therefore assume that this index is
affected by some, as yet unidentified, calibration error (either in our
reductions or in the SSP estimates). Values of log(Age) derived from the
calcium triplet and H$\beta$ index are higher than those given in
Table \ref{agez} by $\sim$ 0.15 while [Fe/H] estimates are lower by
$\sim$0.2 dex. While we observe that these estimates are, qualitatively, in 
line with our results from blue indices, we make no further 
attempt at interpretation of these NIR data.

\section{COMPARISON WITH COMPOSITE MODELS}
\label{comp}
In this section we generate models with composite SFHs and compare the 
predictions with our observations in
an effort to understand galaxy histories. The composite model code was first 
described in Sansom \& Proctor (1998), where we assumed solar abundance 
ratios. Here we extend the models to incorporate 
non-solar abundance ratios. Fourteen elements (listed in Table \ref{abs}) are 
followed self-consistently. These cover most of the heavy element mass 
loss from SNII, SNIa and intermediate mass stars. 
 Our models can allow for inflow of gas (enriched to 
the current level of the ISM or of primordial composition) into a single zone. 
The lowest  metallicity modelled by W94 is for $Z=Z_{\odot}/100$. Therefore
we start the models with 10$^6$ M$_{\odot}$ of gas containing this 
small amount of metals ($Z=2\times 10^{-4}$ 
by mass fraction), assuming solar abundance ratios within this initial metallicity.
Low metallicity stars in our Galaxy have an increasing excess of 
$\alpha$-element abundances (as described by Equation \ref{tea1b}). We allow for 
this in the composite models via the denominator in the exponent of 
Equation \ref{combine} which is varied by up to a factor of 3 for $\alpha$-elements
in the SSP stars. Allowance for $\alpha$-enhanced SSPs at low 
metallicity did not produce large effects in the predicted line-strengths 
in galaxies, which are dominated by higher metallicity stars.
We use V99 SSPs and the T00a method to allow for non-solar abundance 
ratios in the composite models, calling on predictions of line-strength 
changes modelled by TB95. The rate of SNIa is a parameter in our models. 
In the current models we use a rate of $3.8\times 10^{-5}$ SNIa
Gyr$^{-1}$
M$_{\odot}^{-1}$. This is approximately that inferred in our Galaxy (Timmes, 
Woosley \& Weaver 1995) which has an uncertainty of about a factor of two.
Larger SNIa rates will produce stronger iron sensitive lines. We assume 
a Schmidt law with index of one for the star formation rate 
(SFR=C$\times$gas density), and a Salpeter initial mass 
function (IMF). 

\subsection{Primordial Collapse}
\label{pc}
For the primordial model we started with C=4.0 Gyr$^{-1}$,
and a rapid, enriched inflow rate of 10$^7$ M$_{\odot}$ Gyr$^{-1}$, going 
down to a more steady rate of C=0.2 Gyr$^{-1}$ and zero inflow after 0.4 Gyr.
Star formation is followed up to 1.5 Gyr ago. 
In a previous paper (Proctor et al. 2000) we showed that (assuming solar
abundance ratios) rapid collapse and star formation in a primordial 
gas cloud does not produce strong enough metal absorption lines when compared 
with observations of early-type galaxies and spiral bulges. We confirm this 
result here with our non-solar ratio models. This is shown in Figs
\ref{diags} and \ref{IvsHbeta}
where the thickest, short line indicates our predictions for such a primordial
collapse model, for times ranging from 10 to 17 Gyr after the start of star 
formation. These primordial models include a higher rate of 7.6$\times10^{-5}$
SNIa Gyr$^{-1}$ M$_{\odot}^{-1}$. The predicted metal line-strengths are 
too low to account for the 
observed line-strengths. Thus the conventional picture of spheroid formation 
through rapid, early collapse, followed by passive evolution, is ruled
out by the observed strong lines. A similar conclusion was found for 
nearby spheroids by Worthey, Dorman \& Jones (1996) 
and is analogous to the well known G-dwarf problem in our own Galaxy, where 
there are insufficient low metallicity stars compared to predictions of 
closed box models with stars generated with a Salpeter IMF.

\subsection{Models with extended inflow}

 In Proctor et al. (2000) we found that observations of spiral bulges could 
be explained with extended inflow models, with gas inflow over several Gyr 
enriched to the current level of the ISM. This assumed solar abundance ratios. 
Allowing for non-solar ratios we find that 
such models tend to under-produce Fe sensitive features. This is because the 
early feedback from SNII is extremely Mg rich compared to Fe (several 10s of 
times solar ratios - see the SNII models of Woosley and Weaver 1995). To 
produce models which can simultaneously explain Fe and Mg sensitive 
spectral features a delayed burst of star formation seems to be needed to 
allow the ISM to first become enriched with Fe peak products from SNIa. Such 
models are preliminarily explored in the next section.

\subsection{Merger Models}
\label{merger}
With a delayed burst of star formation we begin to be able to produce models 
which can simultaneously explain the strengths of several spectral features. 
A full exploration of SFH parameter space is beyond the scope of this paper
and will be the subject of future work. However, in Figs \ref{diags}
and \ref{IvsHbeta} we 
illustrate predictions of line-strengths for merger models with an
associated burst of star formation several Gyr after the start of 
the SFH (delayed burst models - medium thick short line). 
Present day predictions (at 17 Gyr) for models with a burst ranging from 3 Gyr to 13 
Gyr delay are shown. The parameters used to describe the composite merger 
model shown here are C=4.0 Gyr$^{-1}$ initially,
with low, enriched inflow rate of $ 1\times 10^4$ M$_{\odot}$ Gyr$^{-1}$,
increasing to $10^7$ M$_{\odot}$ Gyr$^{-1}$ during the star-burst.
This rapid star-burst lasts for 0.4 Gyr after which the inflow is set to 
zero, so the remaining gas is rapidly used up in star formation. 
We see that these examples reach the regions populated by early-type 
galaxies, for most Fe and Mg sensitive line-strengths. 
An exception is Fe5406 (but see Fig. \ref{Iwithsig} and Section \ref{early}). 
Interestingly, we find that the models tend to under-predict carbon 
sensitive features ($<$CN$>$, C$_2$4668). This may 
indicate that all the sources of carbon enrichment have not been accounted 
for in our models. Indeed, carbon enhancements in some very low metallicity 
stars in our own Galaxy are hard to explain (Norris, Ryan \& Beers 1997) 
and dredge-up models for the contributions of carbon from intermediate
mass stars are uncertain. 
Stars produced in the delayed burst have lower overall [E/Fe] than the earlier 
star formation, since SNIa 
have had time to accumulate Fe in the ISM. Thus these model predictions 
tend to support the idea of ellipticals forming by mergers of galaxies, 
with enriched gas inflow and enhanced star formation during the merger. 

There are few predictions of detailed galaxy properties from hierarchical 
merger models. However, Kauffmann \shortcite{K96} used a semi-analytic model 
of galaxy formation in both field and cluster environments to make testable
predictions for the ages of various Hubble types in differing
environments. In these models early-type galaxies form by
the merger of two, roughly equal mass, progenitor galaxies, while spiral
galaxies form by accretion of a disc onto a pre-formed elliptical
galaxy. Kauffmann \shortcite{K96} found luminosity-weighted ages for
early types between 
5 and 12.5 Gyr, in good agreement with our findings. Kauffmann \shortcite{K96} 
also found cluster ellipticals to be $\sim$ 4 Gyr older than ellipticals in low 
density environments. This is again consistent with our findings, as the five 
Virgo cluster ellipticals are amongst the oldest in our sample. The positive 
correlation between age and velocity dispersion in elliptical galaxies is 
qualitatively consistent with the relationship proposed by Forbes \& 
Ponman \shortcite{FP99} for $\sigma$ versus merger redshift. Consequently, 
while both primordial collapse and extended inflow models fail to reproduce the main
features of our data without recourse to additional physics (a biased
IMF or Population III stars), the hierarchical 
model of early-type galaxy formation by merger agrees well.

For the bulges of spirals, Kauffmann \shortcite{K96} predicts ages
significantly lower than those in ellipticals, in agreement
with our findings. However, Kauffmann \shortcite{K96} also predicts
a correlation between bulge ages and the luminosity of their discs. 
Inspecting the distribution of Hubble types 
(indicated by solid symbol sizes in Figs \ref{Age} and \ref{Sig})
we find no evidence of a trend in age with spiral Hubble type. Therefore, this
prediction is at odds with our findings. Kauffmann \shortcite{K96} does point 
out, however, that the correlation may be hidden
if there is significant inflow of gas from the disc, perhaps due to the
formation of bars. Many of the bulges in our sample show strong evidence for
kinematic sub-structure and on-going star formation (i.e. emission). Our bulges 
also possess [Mg/Fe] values too low to have been formed in a single primordial burst
(see Fig. \ref{diags}). 
These observations, and the absence of a correlation 
between age and Hubble type, support the idea that disc inflow must play an 
important role in the star formation histories in bulges.

\section{CONCLUSIONS}
\label{concs}
We have derived luminosity-weighted log(Age), iron abundance, abundance
ratios and metallicity (log(Age), [Fe/H], [E/Fe] and [Z/H] respectively) 
in the centres of 32 galaxies, ranging in Hubble type from E to
Sbc. We used 20 indices for which V99 modelled SSPs, plus their
sensitivities to individual elements as tabulated in TB95, to model the
effects of non-solar abundance ratios in these galaxies. We find that
ignoring such enhancements leads to reasonably accurate age and metallicity
estimates, but that using many fewer indices (e.g. 3) leads to larger
errors in derived values (see Table \ref{enablecomps}). By using many 
indices and modelling abundance ratios we are able to probe correlations 
between derived parameters much more accurately.

Our sample  of early-type galaxies spans a wide range of ages (2
to 13 Gyr) with the 5 Virgo cluster ellipticals amongst the oldest. 
These E/S0 galaxies show correlations of both velocity dispersion ($\sigma$) 
and [E/Fe] with age, while [Fe/H] and [Z/H] both show anti-correlations with age. 
[Fe/H], [Z/H] and [E/Fe] show no significant correlation with $\sigma$ over the
small range in $\sigma$ covered by our E/S0 galaxies.
These results are at odds with the predictions of primordial 
collapse models, which predict uniformly-old, early-type galaxies and
increasing [Z/H] with $\sigma$. However, the correlations suggest that the 
main parameter controlling the metal content of our bright, early-type galaxies 
is age.  Our results are consistent with the predictions of hierarchical merger models 
of galaxy formation (e.g. Kauffmann 1996) which predict the observed
E, S0 (and bulge) age sequence. The observed correlation between age and velocity dispersion is 
in qualitative agreement with the relationship proposed by Forbes \& Ponman 
\shortcite{FP99} for merger remnants. The strong correlation
of [E/Fe] with log(Age) and anti-correlation of [Fe/H] with log(Age) are
qualitatively reproduced by our merger models of galaxy formation (Section 
\ref{merger} and Fig. \ref{IvsHbeta}). In our models these trends are
the result of the shortening interval (in which SNIa can produce Fe) 
between the commencement of star formation and the final merger event. Thus we
find several observational results which agree with the predictions of
merger models for early-type galaxies.

The correlations outlined above in turn call for careful interpretation of
Lick index versus $\sigma$ correlations. For instance, high Mg$_2$, high velocity 
dispersion, early-type galaxies are amongst the most metal-\emph{poor} in our sample. 
Any correlation of Mg$_2$ with $\sigma$, may in fact, reflect an increasing 
age and \emph{enhancement} ([E/Fe]) with velocity dispersion
(indicating mass) not, as is usually assumed, a metallicity-mass
relation (see also related comments in the conclusion of Trager et
al. \shortcite{Tea00b}). 

The anti-correlations of [Fe/H] and of [Z/H] with log(Age) could also have
implications for the interpretation of colour-magnitude diagrams of 
these objects. This can be illustrated by the extremely narrow
range of optical colours ($\Delta$(U-V) $\sim$ 0.08 for SSPs ranging in age from 3 to 17 Gyr)
predicted for SSPs that follow the age-[Fe/H] anti-correlation shown
by our early-type galaxy sample. Colour-magnitude correlations have
also been interpreted as metallicity-mass relations on the basis of monolithic
collapse models for the formation of elliptical galaxies (e.g. Kodama
et al. 1998). Our results from line-strengths suggest that merger
models must be considered before this interpretation of the observed
correlations can be relied on.
 
We detect significant differences between early-type galaxies and spiral bulges. 
Es, S0s and bulges in our sample form a continuous overlapping sequence of 
decreasing luminosity-weighted age, with bulges typically 2 Gyr younger than S0s, 
and 5 Gyr younger than Es. This is again in line with the predictions of Kauffmann 
\shortcite{K96}, for galaxies in low density environments. We find no significant 
correlations with age in bulges. However, correlations of [Fe/H] and [Z/H] with $\sigma$
are strong. Thus the main parameter controlling the metal
content of bulges is $\sigma$.
Kauffmann \shortcite{K96} predicts a correlation between bulge-to-disc ratio
and age in spiral
galaxies. We detect no such correlation in our data. However,
Kauffmann \shortcite{K96} also points out that if significant inflow from the disc 
occurs, after a merger, this correlation may be lost. The
relatively low values of [E/Fe] ($\sim$ +0.15) in bulges and the strong
correlations of metals with $\sigma$ are consistent with such a model of
bulge formation. The correlations of [Fe/H] and [Z/H] with $\sigma$ in spiral bulges
means that we interpret the correlations of Lick indices with $\sigma$
found in this work (Figs \ref{Iwithsig} and \ref{AgewithSig} and Table \ref{correls}) as a
metallicity-mass relation, in contrast to our finding for early-type
galaxies.

In conclusion, we have shown that primordial collapse models of galaxy
formation are unable to reproduce the line-strengths observed in the
spheroids of galaxies, while merger models can. 
Derived ages and correlations between derived parameters differ significantly between
early and late-type galaxies, suggesting that, at least at some point in
their evolution, the star formation histories in these objects must have differed
significantly. We therefore contend that the similarities in
morphology and photometric properties in these objects, noted in the
introduction, are the result of the various degeneracies at work rather
than indicating similar formation processes.\\

\noindent {\bf ACKNOWLEDGEMENTS}\\   
The authors acknowledge   the data analysis
facilities provided  by the Starlink  Project which is  run by CCLRC on
behalf of PPARC. In addition, the IRAF software package was used. IRAF
is distributed by the National  Optical Astronomy Observatories, which
is  operated   by AURA,  Inc.,  under cooperative   agreement with the
National  Science  Foundation. This  work  is   based on  observations
made with the William Herschell Telescope, operated on the island of
La Palma, by the Isaac Newton Group in the Spanish Observatorio del
Instituto de Astrofisica, Roque de los Muchachos de Canarias. Travel
funds were provided by PPARC and the University of Central Lancashire.
Thanks go to our colleagues G. Bromage and C. Haines for comments on 
this paper prior to submission. Thanks also to the referee, 
Scott Trager, for careful reading and improvement of this paper.\\

\appendix
\newpage
\section{VELOCITY DISPERSION CORRECTIONS}

\begin{table*}
\caption{Polynomial coefficients relating the velocity
dispersion correction factor (C$_{i}$; see text) to the correction velocity
($\sigma_{C}$).}
\begin{tabular}{|c|c|c|c|c|c|c|}
\hline
Index   & (A)dditive  &  
\multicolumn{4}{c}{{Polynomial Coefficients (C$_{i}$ versus $\sigma_C$ 
in km s$^{-1})$}}& Uncertainty\\
        & or             & $x_{0}$& $x_{1}$   & $x_{2}$   & $x_{3}$ & in correction\\
        &(M)ultiplicative&        &(x 10$^{-3}$)&(x 10$^{-6}$)&(x
10$^{-9}$)&at $\sigma_C$=200 km s$^{-1}$\\
\hline
H$\delta_{A}$&A& 0.0 & -0.058& -5.195 &  3.894& 0.029\AA\\ 
H$\delta_{F}$&A& 0.0 & -0.005&  0.704 & -0.077& 0.019\AA\\
CN$_{1}$     &A& 0.0 &  0.002&  0.097 & -0.081& 0.001mag\\ 
CN$_{2}$     &A& 0.0 &  0.003&  0.225 & -0.201& 0.001mag    \\ 
Ca4227       &M& 1.0 &  0.231&  1.246 & 13.400& 3.3\%     \\     
G4300        &M& 1.0 &  0.021&  0.480 &  0.133& 0.4\%     \\   
H$\gamma_{A}$&A& 0.0 &  0.006&  1.181 & -8.883& 0.053\AA    \\ 
H$\gamma_{F}$&A& 0.0 & -0.079& -2.751 &  4.096& 0.054\AA    \\ 
Fe4383       &M& 1.0 &  0.037&  2.334 & -0.811& 1.7\%     \\ 
Ca4455       &M& 1.0 &  0.086&  3.941 &  1.424& 4.1\%     \\  
Fe4531       &M& 1.0 &  0.033&  1.608 &  0.628& 1.4\%    \\ 
C$_{2}4668$  &M& 1.0 & -0.001&  0.851 &  0.284& 0.5\%     \\ 
H$\beta$     &M& 1.0 &  0.031&  0.213 &  0.384& 1.0\%    \\ 
Fe5015       &M& 1.0 &  0.040&  3.025 & -2.443& 1.1\%       \\     
Mg$_{1}$     &A& 0.0 &  0.002&  0.067 & -0.073& 0.001mag      \\   
Mg$_{2}$     &A& 0.0 &  0.004&  0.035 & -0.008& 0.001mag      \\
Mgb          &M& 1.0 & -0.003&  2.129 &  0.307& 0.4\%       \\ 
Fe5270       &M& 1.0 &  0.024&  3.027 & -2.352& 0.5\%    \\     
Fe5335       &M& 1.0 &  0.059&  4.748 &  1.522& 0.6\%     \\     
Fe5406       &M& 1.0 &  0.050&  3.927 &  3.565& 1.2\%    \\     
Ca1          &M& 1.0 & -0.053&  1.892 &  2.526& 0.1\%  \\
Ca2          &M& 1.0 &  0.004&  1.248 &  2.339& 0.1\%  \\
Ca3          &M& 1.0 &  0.028&  1.592 &  0.909& 0.1\%   \\
MgI          &M& 1.0 & -0.082& 11.286 & -1.443& 0.1\%   \\
\hline
\label{poly}
\end{tabular}
\end{table*}

\begin{figure*}
\vspace{3cm}
Figure available from http://www.star.uclan.ac.uk/\~{}rnp/research.htm
\vspace{20cm}
\caption{Behaviour of indices with broadening.
Stellar correlations are shown as lines. Galaxy data are shown as points.
Only galaxies with total observed broadening below $\sigma_L$ are plotted.}
\label{all_ISig}
\end{figure*}

\subsection{Characterisation of stellar indices with spectral broadening}
\label{char}
For accurate calibration, indices in galaxy spectra require correction
to account for the effects of internal stellar velocity dispersion. 
Consequently, the indices of a sub-sample of Lick calibration stars were 
measured after convolving 
their spectra with Gaussians of a range of widths. 15 stars (spectral types
G8 to K3), which best matched the spectral energy distributions of the
galaxies, were used for the blue indices. In the red, all 24 Lick calibration stars 
were found to match galaxy spectra well. For each index a correction factor 
was calculated at each value of broadening. For most line indices, the correction 
factor was calculated as the ratio (C$_{i}$) of the index value at the Lick 
resolution to that at each value of broadening, i.e:

\begin{equation}
C_{i}=I_{L}/I_{Meas}
\end{equation}

Where I$_{L}$ is the index value at the calibration resolution
and I$_{Meas}$ is the index value in the broadened spectrum.

For molecular band indices (CN1, CN2, Mg$_{1}$ and  Mg$_{2}$) and indices with 
ranges spanning zero (H$\delta$ and H$\gamma$ indices), correction factors
were calculated as the \emph{difference} between the measured index and that at the 
calibration resolution at each value of broadening, i.e:

\begin{equation}
C_{i}=I_{L}-I_{Meas}
\end{equation}

The appropriate correction factor was calculated for each index, at each 
value of broadening, by averaging values from the stellar sub-samples. A 
polynomial fit of order 3 was found such that:

\begin{equation}
C_{i}=x_{0}+x_{1}\sigma_{C}+x_{2}\sigma_{C}^{2}+x_{3}\sigma_{C}^{3}.
\end{equation}

Where $\sigma_{C}$ is the width of the Gaussian (in km s$^{-1}$)
convolved with the stellar spectrum at $\sigma_{L}$.
Behaviours of the indices of the stellar sample with broadening
above the Lick resolution are shown in Fig. \ref{all_ISig} 
as lines. These are in good agreement with the behaviours of 
indices measured by previous authors (e.g. Kuntschner 2000). 
The nature of the applied velocity dispersion corrections, i.e.
whether C$_{i}$ should be  multiplied by the raw index value 
or added to the raw value are indicated in the plots and
summarised in column 2 of Table \ref{poly}. This table also gives a typical 
uncertainty in the correction factor for each index, estimated as the RMS 
scatter in the stellar values about the mean C$_{i}$ at $\sigma_{C}$=200 
kms$^{-1}$. This corresponds to the  appropriate value for a high velocity 
dispersion elliptical galaxy. When correction factors are used 
the scatter in the stellar data at the value of broadening appropriate
to the galaxy was included in the statistical errors.

\subsection{Application to galaxies}
To gain confidence in these polynomials, 
galaxy spectra were also broadened by convolution with Gaussians of a range of 
widths. The behaviour of the galaxies 
was then compared to that of the stellar sub-sample (Fig. \ref{all_ISig}). 
For most indices the galaxy data lie within the scatter of the stellar
data. However, for a few 
indices, large scatter and/or small systematic differences in the 
behaviour of galaxies compared to the stellar data were observed. These 
indices (most noticeably H$\delta$, Ca4227 and
Ca4455) are among those affected by the poor removal of the dichroic
response. The scatter in the stellar data has been allowed for in our 
errors as described in Section \ref{flux}.
However, to test the possible impact of differences 
in behaviour between our samples of stars and galaxies, two velocity 
dispersion correction techniques were tested.\\

\noindent \emph{Technique 1.}\\
The first technique (which is generally used by other authors) is to 
broaden all galaxy spectra by a Gaussian of a width $\sigma_{B}$, given by 
Equation \ref{sigB}, prior to measurement of indices. The resultant spectrum 
of a galaxy with velocity dispersion $\sigma_{V}$, then has a total broadening 
($\sigma_{G}$) given by:

\begin{equation}
\sigma_{G}^2 = \sigma_{I}^2 + \sigma_{V}^2+\sigma_{B}^2
\end{equation}

\begin{equation}
\hspace{5mm}= \sigma_{L}^2 + \sigma_{V}^2
\end{equation}

As the correction factors are dependent on the \emph{excess} broadening
of the galaxy with respect to $\sigma_{L}$
the value $\sigma_{C}$  = $\sigma_{V}$ is entered into the polynomials 
in Table \ref{poly} to obtain the correction factor C$_{i}$.\\

\noindent \emph{Technique 2.}\\
In the second technique, if the total broadening of the galaxy 
$(\sigma_I^2$+$\sigma_V^2)^{1/2}$ is less than
the target calibration resolution ($\sigma_L$), the galaxy spectrum 
is broadened up to the Lick resolution by convolution with a Gaussian 
of width $\sigma_{B}$ given by:

\begin{equation}
\sigma_{B}^2=\sigma_{L}^2-\sigma_{I}^2-\sigma_{V}^2\\ 
\end{equation}

Consequently, as the spectrum has been broadened to the calibration 
resolution, the final index values can be measured directly from the
broadened galaxy spectrum with no need for correction.\\

If, on the other hand, the total galaxy broadening is greater 
than $\sigma_L$, then the spectra are left 
un-broadened and $\sigma_{C}$=($\sigma_{I}^2 +
\sigma_{V}^2-\sigma_{L}^2)^{1/2}$ is substituted into the 
polynomials to calculate the appropriate value of C$_{i}$.\\

It should be noted that, for some indices (e.g. H$\delta$), the second 
technique involves no use of the polynomials as 
all galaxies in our sample have $\sigma_{L}^{2}>\sigma_{I}^{2}+\sigma_{V}^{2}$.
The two techniques agree well over our whole velocity
dispersion range. This is true for all indices measured. 
Comparisons of the results from the two techniques can be found in
Proctor (PhD thesis - in preparation). In this paper
the second technique is used since it involves smaller corrections,
thus minimising the effects of systematic differences in the
behaviour of the stellar and galaxy spectra with broadening.


\begin{thebibliography}{99}

\bibitem[\protect\citename{Balcells \& Peletier }1994]{BP94}
    Balcells M., Peletier R.F., 1994, AJ, 107, 135

\bibitem[\protect\citename{Barnes \& Hernquist }1996]{BH96}
    Barnes J.E., Hernquist L., 1996, ApJ, 471, 115

\bibitem[\protect\citename{Bender et al. }1993]{BBF93}
    Bender R., Burstein D., Faber S.M., 1993, ApJ, 411, 153

\bibitem[\protect\citename{Bender et al. }1998]{Ben98}
    Bender R., Saglia R.P., Ziegler B., Belloni P., Greggio L.,
    Hopp U., Bruzual G., 1998, ApJ, 493, 529

\bibitem[\protect\citename{Bernardi et al. }1998]{Ber98}
    Bernardi M., Renzini A., da Costa L.N., Wegner G., Alonso M.V., 
    Pellegrini P.S., Rit\'e C., Willmer C.N.A., 1998, ApJ, 508, L143

\bibitem[\protect\citename{Bertelli et al. }1994]{BBCNF}
    Bertelli G., Bressan A., Chiosi C., Fagotto F., Nasi E., 1994,
    A\&AS, 106, 275

\bibitem[\protect\citename{Carlberg }1984]{C84}
    Carlberg R.G., 1984, ApJ, 286, 403

\bibitem[\protect\citename{Ciotti et al. }1991]{Cea91}
    Ciotti L., D'Ercole A., Pellegrini S., Renzini A., 1991, ApJ, 376, 380

\bibitem[\protect\citename{Concannon et al. }2000]{Cea00}
    Concannon K.D., Rose J.A., Caldwell N., 2000, ApJ, 536, L19

\bibitem[\protect\citename{Cox }2000]{C00}
    Cox A.N., 2000, Allen's Astrophysical Quantities 4th ed. p29, AIP
    Press, Springer-Verlag, New York
    
\bibitem[\protect\citename{Davies, Sadler \& Peletier }1993]{DSP93}
    Davies R.L., Sadler E.M., Peletier R.F., 1993, MNRAS, 262, 650

\bibitem[\protect\citename{de Vaucouleurs et al. }1991]{RC3}
    de Vaucouleurs G., de Vaucouleurs A., Corwin H.G.,
    Buta R.J., Paturel G., Fouqu\'e P., 1991, Third Reference
    Catalogue of Bright Galaxies. Springer-Verlag, New York ({\bf RC3})
       
\bibitem[\protect\citename{Diaz et al. }1989]{DTT89}
    Diaz A.I., Terlevich E., Terlevich R., 1989, MNRAS, 239, 325

\bibitem[\protect\citename{Edvardsson et al. }1993]{Eea93}
    Edvardsson B., Andersen J., Gustafsson B., Lambert D.L.,
    Nissen P.E., Tomkin J., 1993, A\&A, 275, 101

\bibitem[\protect\citename{Faber }1973]{F73}
    Faber S.M., 1973, ApJ, 179, 731

\bibitem[\protect\citename{Faber et al. }1985]{FFBG85}
    Faber S.M., Friel E.D., Burstein D., Gaskell C.M.,
    1985, ApJS, 57, 711

\bibitem[\protect\citename{Feltzing \& Gustafsson. }1998]{FG98}
    Feltzing S., Gustafsson B., 1998, A\&AS, 129, 237

\bibitem[\protect\citename{Fisher Franx \& Illingworth. }1996]{FFI96}
    Fisher D., Franx M., Illingwoth G., 1996, ApJ, 459, 110

\bibitem[\protect\citename{Forbes \& Ponman }1999]{FP99}
    Forbes D.A., Ponman T.J., 1999, MNRAS, 309, 623

\bibitem[\protect\citename{Gonz\'{a}lez }1993]{G93}
    Gonz\'{a}lez J.J., 1993, PhD Thesis, Univ. California.

\bibitem[\protect\citename{Gorgas et al. }1993]{Gea93}
    Gorgas J., Faber S.M., Burstein D., Gonz\'{a}lez J.J.,
    Courteau S., Prosser C. 1993, ApJS, 86, 153

\bibitem[\protect\citename{Gorgas et al. }1997]{Gea97}
    Gorgas J., Pedraz S., Guzm\'{a}n R., Cardiel N.,
    Gonz\'{a}lez J.J., 1997, ApJ, 481, L19
        
\bibitem[\protect\citename{Goudfrooij \& Emsellem }1996]{GE96}
    Goudfrooij P., Emsellem E., 1996, A\&A, 306, L45

\bibitem[\protect\citename{Goudfrooij, Gorgas  \& Jablonka }1999]{GGJ99}
    Goudfrooij P., Gorgas J., Jablonka P., 1999, Ap\&SS, 269, 109

\bibitem[\protect\citename{Gustafsson, Kj{\ae}rgaard, Andersen }1974]{GKA74}
    Gustafsson B., Kj{\ae}rgaard P., Andersen S., 1974, A\&A, 34, 99
     
\bibitem[\protect\citename{Greggio }1997]{G97}
     Greggio L., 1997, MNRAS, 285,151

\bibitem[\protect\citename{Hansen \& Kj{\ae}rgaard }1971]{HK71}
     Hansen L.,  Kj{\ae}rgaard P., 1971, A\&A 15,123

\bibitem[\protect\citename{Henry \& Worthey }1999]{HW99}
    Henry R.B.C., Worthey G., 1999, PASP, 111, 919

\bibitem[\protect\citename{Idiart \& Thevenin }2000]{IT00}
    Idiart T., Th\'evenin F., 2000, ApJ, 541, 2071

\bibitem[\protect\citename{Idiart, de Freitas Pacheco \& Costa}1996]{IFC96}
    Idiart T., de Freitas Pacheco J.A., Costa R.D.D.,1996, AJ, 112, 2541

\bibitem[\protect\citename{Jablonka, Martin \& Arimoto }1996]{JMA96}
    Jablonka P., Martin P., Aromoto N., 1996, AJ, 112, 1415

\bibitem[\protect\citename{J{\o}rgensen }1997]{J97}
    J{\o}rgensen I., 1997, MNRAS, 288, 161

\bibitem[\protect\citename{J{\o}rgensen }1999]{J99}
    J{\o}rgensen I., 1999, MNRAS, 306, 607

\bibitem[\protect\citename{Kauffmann }1996]{K96}
    Kauffmann G., 1996, MNRAS, 281, 487

\bibitem[\protect\citename{Kauffmann White \& Guiderdoni }1993]{KWG93}
    Kauffmann G., White S.D.M., Guiderdoni B., 1993, MNRAS, 264, 201

\bibitem[\protect\citename{Khosroshahi Wadadekar \& Kembhavi }2000]{KWK00}
    Khosroshahi H.G., Wadadekar Y., Kembhavi A., 2000, ApJ, 533, 162

\bibitem[\protect\citename{Kodama et al. }1998]{Kea98}
    Kodama T., Arimoto N., Barger A.J., Arag\'{o}n-Salamanca A., 1998,
    A\&A, 334, 99

\bibitem[\protect\citename{Kuntschner }2000]{K00}
    Kuntschner H., 2000, MNRAS, 315, 184 ({\bf K00})

\bibitem[\protect\citename{Kuntschner \& Davies }1998]{KD98}
    Kuntschner H., Davies R.L., 1998, MNRAS, 295, L29 

\bibitem[\protect\citename{Kuntschner et al. }2001]{Kea01}
    Kuntschner H., Lucey J.R., Smith R.J., Hudson M.J.,
    Davies R.L., 2001, MNRAS, 323, 615
    
\bibitem[\protect\citename{Lebreton et al. }1999]{Lea99}
    Lebreton Y., Perrin M.-N., Cayrel R., Baglin A., Fernandes J.,
    1999, A\&A, 350, 587 

\bibitem[\protect\citename{Norris, Ryan \& Beers }1997]{NRB97}
    Norris J.E., Ryan S.G., Beers T.C., 1997, ApJ 488, 350

\bibitem[\protect\citename{O'Connell }1976]{OC76}
    O'Connell R.W., 1976, ApJ, 206, 370

\bibitem[\protect\citename{Osterbrock }1989]{O89}
    Osterbrock D.E., 1989, Astrophysics of Gaseous Nebulae and Active
    Galactic Nuclei, University Science Books, Mill Valley
    
\bibitem[\protect\citename{Prugniel \& Simien }1997]{PS97}
    Prugniel P., Simien F., 1997, Web site, 
    ftp://ftp-obs.univ-lyon1.fr/pub/galaxies/data/kinematics/sigma97.txt
    
\bibitem[\protect\citename{Proctor, Sansom \& Reid }2000]{PSR00}
    Proctor R.N., Sansom A.E., Reid I.N., 2000, MNRAS, 311, 37

\bibitem[\protect\citename{Ryan Norris \& Bessell }1991]{Rea91}
    Ryan S.G., Norris J.E., Bessell M.S., 1991, AJ, 102, 303

\bibitem[\protect\citename{Sansom \& Proctor }1998]{SP98}
    Sansom A.E., Proctor R.N., 1998, MNRAS, 297, 953

\bibitem[\protect\citename{Salasnich et al. }2000]{Sea00}
    Salasnich B., Girardi L., Weiss A., Chiosi C., 2000, A\&A, 361, 1023

\bibitem[\protect\citename{Salaris \& Weiss }1998]{SW98}
    Salaris M., Weiss A., 1998, A\&A, 335, 943

\bibitem[\protect\citename{Tantalo, et al. }1998]{TCB98}
    Tantalo R., Chiosi C., Bressan A., 1998, A\&A, 333, 419

\bibitem[\protect\citename{Timmes, Woosley \& Weaver }1995]{TWW95}
    Timmes F.X., Woosley S.E., Weaver T.A., 1995, ApJS, 98, 617

\bibitem[\protect\citename{Trager }1998]{T98}
    Trager S.C., 1998, PhD Thesis, University of California 

\bibitem[\protect\citename{Trager et al. }1998]{Tea98}
    Trager S.C., Worthey G., Faber S.M., Burstein D., Gonz\'{a}lez
    J.J.,1998, ApJS 116, 1

\bibitem[\protect\citename{Trager et al. }2000a]{Tea00a}
    Trager S.C., Faber S.M., Worthey G.,Gonz\'{a}lez J.J.,2000a,
    AJ, 119, 1645 ({\bf T00a})

\bibitem[\protect\citename{Trager et al. }2000b]{Tea00b}
    Trager S.C., Faber S.M., Worthey G.,Gonz\'{a}lez J.J.,2000b, AJ, 120, 165

\bibitem[\protect\citename{Tripicco \& Bell }1995]{TB95}
    Tripicco M.J., Bell R.A., 1995, AJ, 110, 3035 {\bf (TB95)}

\bibitem[\protect\citename{Tully }1988]{T88}
    Tully R.B., 1988, Nearby Galaxies Catalogue, Camb. Univ. Press,
    Cambridge
   
\bibitem[\protect\citename{VandenBerg et al. }2000]{Vea00}
    VandenBerg D.A., Swenson F.J., Rogers F.J., Iglesias C.A.,
    Alexander D.R., 2000, ApJ, 532, 430

\bibitem[\protect\citename{Vazdekis }1999a]{V99a}
    Vazdekis A., 1999a, ApJ, 513, 224

\bibitem[\protect\citename{Vazdekis }1999b]{V99b}
    Vazdekis A., 1999b,\\
    http://www.iac.es/galeria/vazdekis/col\_lick.html {\bf (V99)}

\bibitem[\protect\citename{Vazdekis et al. }1996]{Vea96}
    Vazdekis A., Casuso E., Peletier R. F., Beckman J. E., 1996, ApJS,
    106, 307 {\bf (V96)}

\bibitem[\protect\citename{Vazdekis et al. }1997]{Vea97}
    Vazdekis A., Peletier R. F., Beckman J. E., Casuso E., 1997, ApJS,
    111, 203

\bibitem[\protect\citename{Woosley \& Weaver }1995]{WW95}
    Woosley S.E., Weaver T.A., 1995, ApJS, 101, 181

\bibitem[\protect\citename{Worthey }1994]{W94}
    Worthey G., 1994, ApJS, 95, 107 {\bf (W94)}

\bibitem[\protect\citename{Worthey }1998]{W98}
    Worthey G., 1998, PASP, 110, 888 

\bibitem[\protect\citename{Worthey \& Ottaviani }1997]{WO97}
    Worthey G., Ottaviani D.L., 1997, ApJS, 111, 377

\bibitem[\protect\citename{Worthey, Faber \& Gonz\'{a}lez }1992]{WFG92}
    Worthey G., Faber S.M., Gonz\'{a}lez J.J., 1992, ApJ, 398, 69

\bibitem[\protect\citename{Worthey, Faber, Gonz\'{a}lez \& Burstein}1994]{Wea94}
    Worthey G., Faber S.M., Gonz\'{a}lez J.J., Burstein D., 1994,
    ApJS, 94, 687
    
\bibitem[\protect\citename{Worthey, Dorman \& Jones}1996]{WDJ96}
    Worthey G., Dorman B., Jones L.A., 1996, AJ, 112, 948


\label{lastpage}
\end{thebibliography}
\end{document}